\documentclass[10pt,journal,compsoc]{IEEEtran}
\usepackage{algorithm}
\usepackage{algorithmic}
\usepackage{amsmath}
\usepackage{amssymb}
\usepackage{amsthm}
\usepackage{subfig}
\usepackage{threeparttable}
\usepackage{footnote}
\makesavenoteenv{tabular}
\usepackage{tabularx}

\usepackage{progressbar}
\usepackage{makecell}
\usepackage{tabularx}
\usepackage{multicol}
\usepackage{multirow}
\usepackage{hyperref}

\usepackage{xparse}

\newtheorem{assumption}{Assumption}
\newtheorem{lemma}{Lemma}

\newtheorem{theorem}{Theorem}
\newtheorem{corollary}{Corollary}[theorem]

\usepackage{xcolor}
\usepackage{color}

\newcommand{\gxs}[1]{{\mathbf{x}}_{#1}}
\newcommand{\gx}[1]{\tilde{\mathbf{x}}_{#1}}
\newcommand{\lx}[3]{\mathbf{x}_{#2, #3}^{(#1)}}
\newcommand{\optx}{\mathbf{x}_*}
\newcommand{\optxs}{\mathbf{x}^*}
\NewDocumentCommand\x{ggg}{%
    \IfNoValueTF{#1}{\optx}{\IfNoValueTF{#2}{\gx{#1}}{\lx{#1}{#2}{#3}}}
}

\newcommand{\w}{\omega}
\newcommand{\E}{\mathbb{E}}
\newcommand{\bracket}[1]{\left(#1\right)}
\newcommand{\norm}[1]{\left\|#1\right\|_2^2}
\newcommand{\innerproduct}[2]{\left\langle#1, #2\right\rangle}

\newcommand{\gradone}[1]{\nabla F\left(\gx{#1}\right)}
\newcommand{\gradtwo}[2]{\nabla F_{#1}\left(\gx{#2}\right)}
\newcommand{\gradtwos}[2]{\nabla F_{#1}\left(\gxs{#2}\right)}
\newcommand{\gradthree}[3]{\nabla F_{#1}\left(\x{#1}{#2}{#3}\right)}
\NewDocumentCommand\grad{mgg}{%
    \IfNoValueTF{#2}{\gradone{#1}}{\IfNoValueTF{#3}{\gradtwo{#1}{#2}}{\gradthree{#1}{#2}{#3}}}
}

\newcommand{\revise}[1]{\textcolor{black}{#1}}

\begin{document}
\title{From Deterioration to Acceleration: A Calibration Approach to Rehabilitating Step Asynchronism in Federated Optimization}

\author{Feijie Wu, Song Guo, \IEEEmembership{Fellow, IEEE,} Haozhao Wang, Haobo Zhang, Zhihao Qu, \IEEEmembership{Member, IEEE,} Jie Zhang, \IEEEmembership{Member, IEEE,} Ziming Liu
\IEEEcompsocitemizethanks{
\IEEEcompsocthanksitem Song Guo and Zhihao Qu are the corresponding authors.
\IEEEcompsocthanksitem Feijie Wu, Song Guo, Jie Zhang and Ziming Liu are with the Department of Computing, The Hong Kong Polytechnic University, Hong Kong (email: \{harli.wu, jieaa.zhang, ziming.liu\}@connect.polyu.hk, song.guo@polyu.edu.hk) 
\IEEEcompsocthanksitem Haozhao Wang is with the School of Computer Science and Technology, Huazhong University of Science and Technology, Wuhan, China (e-mail: hz\_wang@hust.edu.cn).
\IEEEcompsocthanksitem Haobo Zhang is with Computer Science and Engineering department, Michigan State University, East Lansing, MI, USA (email: zhan2060@msu.edu). 
\IEEEcompsocthanksitem Zhihao Qu is with the School of Computer and Information, Hohai University, Nanjing, China (e-mail: quzhihao@hhu.edu.cn).
}

}

% \markboth{IEEE TRANSACTIONS ON PARALLEL AND DISTRIBUTED SYSTEMS,~Vol.~xx, No.~xx, February 2023}%
% {Wu \MakeLowercase{\textit{et al.}}: Bare Demo of IEEEtran.cls for Computer Society Journals}

\IEEEtitleabstractindextext{%
\begin{abstract}
In the setting of federated optimization, where a global model is aggregated periodically, step asynchronism occurs when participants conduct model training by efficiently utilizing their computational resources. It is well acknowledged that step asynchronism leads to objective inconsistency under non-i.i.d. data, which degrades the model's accuracy. To address this issue, we propose a new algorithm \texttt{FedaGrac}, which calibrates the local direction to a predictive global orientation. Taking advantage of the estimated orientation, we guarantee that the aggregated model does not excessively deviate from the global optimum while fully utilizing the local updates of faster nodes. We theoretically prove that \texttt{FedaGrac} holds an improved order of convergence rate than the state-of-the-art approaches and eliminates the negative effect of step asynchronism. Empirical results show that our algorithm accelerates the training and enhances the final accuracy. 
\end{abstract}

}

\maketitle

\IEEEdisplaynontitleabstractindextext

% \fi
\IEEEpeerreviewmaketitle

\IEEEraisesectionheading{\section{Introduction}\label{sec:introduction}}

\IEEEPARstart{F}{ederated} learning (FL) is thriving as a promising paradigm that refrains the leakage of users' data, including raw information and labels distribution. With the rapid development of FL techniques over the past few years, a wide range of applications for computer vision \cite{liu2020fedvision, yu2019federated} and natural language processing \cite{liu2020federated, wu2020fedmed, chen2019federated} have deployed over a large set of edge devices (e.g., smartphones and tablets). Conventionally, clients perform a fixed number of local \revise{stochastic gradient descent (SGD)} steps in each round; then, the server aggregates the updated models and finally acquires and distributes the global one to all clients \cite{mcmahan2017communication, li2019convergence}. FedAvg follows the preceding procedure and has been proven to be a promising solution to data heterogeneity. 

\begin{table}[t]
% \shiqi{Only takes 1 round to reach 80\%? I need some explainations.}
  \caption{The number of communication rounds to reach the test accuracy of 80\% under Logistic Regression (LR) and 2-layer CNN on Fashion-MNIST with various settings when 10 devices participate in FedAvg. 
  The number of local updates is 100 without step asynchronism, while \revise{under step asynchronism, clients perform at least 100 local updates}.
  The learning rates set for LR and 2-layer CNN are 0.001 and 0.03, respectively, which are also applied to Table \ref{table:util_acc}, Figure \ref{fig:various_lam}, and Figure \ref{fig:fedagrac_evaluation}. }
% \shiqi{No need to specify learning rates? The current caption is too long} And without special annotation, these settings are also applied to Table \ref{table:util_acc}, Figure \ref{fig:various_lam}, and Figure \ref{fig:fedagrac_evaluation}.
\centering
\renewcommand{\arraystretch}{1.2}
  \begin{tabular}{lclcl}
  \Xhline{3\arrayrulewidth}
  FedAvg with & \multicolumn{2}{c}{LR} & \multicolumn{2}{c}{2-layer CNN}  \\\hline
  neither & \makecell{\progressbar[width=1.6cm, ticksheight=0, linecolor=black, filledcolor=black]{0.01}} & 2 & \makecell{  \progressbar[width=1.6cm, ticksheight=0, linecolor=black, filledcolor=black]{0.10}} &20 \\
  step async & \makecell{\progressbar[width=1.6cm, ticksheight=0, linecolor=black, filledcolor=black]{0.005}} & 1 & \makecell{  \progressbar[width=1.6cm, ticksheight=0, linecolor=black, filledcolor=black]{0.04}} & 8 \\
  non-i.i.d. & \makecell{\progressbar[width=1.6cm, ticksheight=0, linecolor=black, filledcolor=black]{0.45}} & 91 & \makecell{  \progressbar[width=1.6cm, ticksheight=0, linecolor=black, filledcolor=black]{0.76}} & 265 \\
  both & \makecell{\progressbar[width=1.6cm, ticksheight=0, linecolor=black, filledcolor=black]{5}} & 1K+ & \makecell{  \progressbar[width=1.6cm, ticksheight=0, linecolor=black, filledcolor=black]{0.96}}&339 \\\Xhline{3\arrayrulewidth}
  \end{tabular}
  \label{table:fedavg}
\end{table}

With an increasing number of nodes participating in the training, the traditional framework becomes infeasible because the computation capacities are substantially diverse among devices~\cite{chai2019towards}. \revise{A practical framework allows clients to update the local model via a flexible number of local SGD steps in each round according to its available resource capacity. And we define such a procedure as \textit{step asynchronism} (see Figure \ref{fig:pretest} for visualized demonstration). } To comprehensively understand the training performance of the traditional algorithm FedAvg, Table \ref{table:fedavg} compares the results in terms of test accuracy in two situations -- step asynchronism and data heterogeneity. This experiment is under convex (i.e., logistic regression) and non-convex (i.e., 2-layer CNN) objectives using a public dataset Fashion-MNIST \cite{xiao2017fashionmnist}. Performance deterioration is noticeable, especially in the logistic regression model the desired test accuracy cannot be reached. 
% that cannot reach the designated test accuracy. 
% \shiqi{What is step asynchronism? Not clear to me. Better to explain with figures.}

% \begin{figure}[t]
% \flushleft
% % \subfloat[Utilization under LR] {\includegraphics[width=.24\textwidth]{Figures/utilization_accuracy/LR_util.pdf} \label{fig:util_lr}}
% % \subfloat[Accuracy under LR] {\includegraphics[width=.24\textwidth]{Figures/utilization_accuracy/LR_acc.pdf} \label{fig:acc_lr}}
% \subfloat[Utilization under CNN] {\includegraphics[width=.24\textwidth]{Figures/utilization_accuracy/cnn_util.pdf}\label{fig:util_cnn}}
% \subfloat[Accuracy under CNN] {\includegraphics[width=.24\textwidth]{Figures/utilization_accuracy/cnn_acc.pdf}\label{fig:acc_cnn}}
% \caption[t]{Utilization and test accuracy under the setting that one Nvidia GTX 3080Ti and nine other devices are shown on the $x$-axis. Here we evaluate Fashion-MNIST with non-convex objectives (i.e., 2-layer CNN). Utilization is measured by the fraction of available local updates w.r.t. maximum local updates when the model achieves the accuracy of 70\%. The red dot line in Figure \ref{fig:acc_cnn} indicates the test accuracy after 200 communication rounds by employing FedAvg with ten Nvidia GTX 3080Ti GPUs.}
% \label{fig:util_acc}
% \end{figure}

\begin{table}[!t]
\centering
\renewcommand\arraystretch{1.2}
\caption{Utilization and test accuracy under the setting that one Nvidia GTX 3080Ti and nine other devices are shown in the first column. Here we evaluate Fashion-MNIST with non-convex objectives (i.e., 2-layer CNN), and the data distribution among clients is heterogeneous. }
\begin{threeparttable}[t]
\newcolumntype{Y}{>{\centering\arraybackslash}X}
% \resizebox{.47\textwidth}{!}{ 
\begin{tabularx}{.47\textwidth}{YYYY}
\Xhline{1.5pt}
 Device & Method & \begin{tabular}[c]{@{}c@{}}Utilization\tnote{1} \\ (Rounds)\end{tabular} & Accuracy(\%)\tnote{2} \\ \Xhline{1pt}

\multirow{2}{*}{\begin{tabular}[c]{@{}c@{}}Raspberry\\ Pi 4\end{tabular}}   
& FedNova  &  25\% (49)   &  66.18  \\ \cline{2-4} 
& \texttt{FedaGrac}  &  \textbf{100\% (25)} &  \textbf{72.07}  \\ \Xhline{1pt} 

\multirow{2}{*}{\begin{tabular}[c]{@{}c@{}}Nvidia\\ Jetson Nano\end{tabular}}   
& FedNova  &  50\% (50)   &  64.39  \\ \cline{2-4} 
& \texttt{FedaGrac}  &  \textbf{100\% (21)} &  \textbf{72.93}  \\ \Xhline{1pt} 

\multirow{2}{*}{\begin{tabular}[c]{@{}c@{}}Nvidia GTX\\ 1080 Ti\end{tabular}}   
& FedNova  &  100\% (40) &  69.77  \\ \cline{2-4} 
& \texttt{FedaGrac}  &  \textbf{100\% (30)} &  \textbf{72.00}  \\ \Xhline{1pt} 

\multirow{2}{*}{\begin{tabular}[c]{@{}c@{}}Nvidia GTX\\ 2080 Ti\end{tabular}}   
& FedNova  &  \textbf{100\% (29)} &  72.11  \\ \cline{2-4} 
& \texttt{FedaGrac}  &  100\% (36) &  \textbf{72.13}  \\ \Xhline{1.5pt} 
\end{tabularx}
% }
\begin{tablenotes}[flushleft]
\setlength\labelsep{0pt}
\item[1] Utilization means the maximum computation capacity of Nvidia GTX 3080Ti to achieve the test accuracy of 60\% in the first 50 rounds. We obtain the value by tuning the resource usage from 100\% and looping a deduction of 5\%. Rounds quantify when the approach achieves 60\% test accuracy. 
\item[2] Given the utilization, we measure the test accuracy after 100 rounds. 
\end{tablenotes}
\end{threeparttable}
\label{table:util_acc}
\end{table}

A previous study \cite{wang2020tackling} owes the performance deterioration to \textit{objective inconsistency}, where the FL training converges to a stationary point that mismatches the optimal solution. In order to alleviate the issue, Wang et al. \cite{wang2020tackling} introduce FedNova, a normalization approach that averages the normalized local gradients and accordingly updates the global model at the server. However, in \revise{Table \ref{table:util_acc}}, we empirically disclose that FedNova cannot fully utilize the computational resources of the powerful node under heterogeneous environments, which explicitly limits the number of local updates for the faster node. 
% \shiqi{What phenomenon? Performance deterioration?}

% Since we do not know the correct global direction, it can be estimated according to the clients’ local updates
In this paper, we propose a method named \texttt{FedaGrac} to conquer the objective inconsistency challenge under a highly imbalanced computational setting in FL. The core idea of our proposed algorithm is to calibrate each local update according to the global update orientation. \revise{Although the correct global direction is not known, it can be estimated based on the clients' local updates: If a client performs the local updates very fast, then the client will transmit the first gradient; otherwise, the averaged gradient.}  By this means, the negative effect of deviation on the convergence can be significantly mitigated. We conduct preliminary experiments and depict the comparison between our proposed algorithm and FedNova \cite{wang2020tackling} in terms of resource utilization and test accuracy in \revise{Table \ref{table:util_acc}}. In all cases, \texttt{FedaGrac} not only fully utilizes the computational resources, but also achieves a better accuracy than FedNova (see \revise{Table \ref{table:util_acc}}). 

Our key contributions to this work are listed as follows: 

\begin{enumerate}
  \item To explore the factors that lead to performance deterioration, we analyze the convergence property under strongly-convex objectives. The theoretical result indicates that the expected loss never reaches the optimal one when both data heterogeneity and step asynchronism exist. In other words, a constant number of local updates eliminates the negative effect of data distribution differentiation, while step asynchronism magnifies the drawback of data heterogeneity. 
  \item We design a novel method named \texttt{FedaGrac} to address the problem of objective inconsistency via predictive gradient calibration, which makes the direction of each local update close to the direction towards the global optimum. For the first time, our algorithm can jointly address statistical heterogeneity and computation heterogeneity at a time.
  % \revise{without prior knowledge}.  
  \item We establish the convergence rate of \texttt{FedaGrac}. Under non-convex objectives, the algorithm achieves a convergence rate of  $O\left(1/\sqrt{MT\bar{K}}\right)$, where $M$ and $T$ represent the number of clients and communication rounds, respectively, and $\bar{K}$ indicates the weighted averaged number of local updates. This convergence rate is also achieved by FedNova only under the condition that $K_{\max} / K_{\min} = O(M)$, where $K_{\max}$ and $K_{\min}$ separately refer to the maximum and minimum number of local updates \cite{wang2020tackling}. Otherwise, the actual convergence rate of FedNova should be $O\left(\sqrt{\bar{K}/MT}\right)$. Apparently, our algorithm can achieve a faster convergence rate by a factor up to $O(\bar{K})$. 
%   Also, to the best of our knowledge, it is the first work that analyzes the convergence rate of strongly-convex objectives under step asynchronism and non-i.i.d. data.
  \item We conduct extensive experiments to compare the proposed \texttt{FedaGrac} with typical and latest works such as SCAFFOLD \cite{karimireddy2020scaffold} and FedNova \cite{wang2020tackling}. In terms of convergence rate, \texttt{FedaGrac} achieves higher convergence efficiency compared to FedAvg and FedNova, especially in scenarios with high heterogeneity. For example, in terms of test accuracy, our algorithm can always preserve convergence while SCAFFOLD and FedNova cannot in some cases. 
  % \shiqi{Delete this example. Not clear.}
\end{enumerate}

The rest of this paper is organized as follows. First, Section \ref{sec:related_work} provides related work and background knowledge of distributed SGD and existing solutions to heterogeneous training. Next, we state preliminaries and problem formulation for the heterogeneous Federated Learning in Section \ref{sec:preliminary}. Then, in Section \ref{sec:FedaGrac}, we design a novel algorithm \texttt{FedaGrac} to solve the problem. In Section~\ref{sec:theory}, we analyze its convergence property. After that, we present our experimental results to evaluate our method in Section \ref{sec:experiment}. Finally, Section \ref{sec:conclusion} concludes the paper.

\section{Related Work} \label{sec:related_work}

% introduce the recent work on data heterogeneity (SCAFFOLD) and step asynchronism (FedNova) 

% For solving SGD in a heterogeneous environment, the existing methods can be categorized into two types: designing asynchronous aggregation scheme and setting adaptive local iterations. 

\textbf{Federated learning.} Frequently, edge devices such as smartphones possess abundant data, which are highly sensitive but useful to the model training \cite{guo2022edge, han2020accelerating, lim2021decentralized, wang2021comprehensive}. To utilize these data, FL is conceived to search for a generalized model \cite{wang2021losp, qu2021partial, zhang2022adaptive} or personalized models \cite{zhang2021parameterized, t2020personalized, tang2021personalized} while safeguarding the data privacy \cite{mcmahan2017communication, konevcny2015federated}. Apparently, the data are heterogeneous among clients because there are no predefined rules for the data distribution for each client. Besides, due to the hardware differences among devices, the computational capabilities are various. In this section, we briefly investigate the flaws raised by data heterogeneity and computation heterogeneity and review the existing work to tackle these two issues. 

% Client sampling (computational heterogeneity) to reduce the waiting time \cite{huang2020efficiency, deng2021auction, zhou2020petrel, wu2020accelerating}

% communication reduction \cite{zhou2021communication, wu2020fedscr, wang2021error, wu2020convergence} 

% How to improve the performance of FedAvg is burgeoning as a heated topic over the past few years. This section categorizes these methods into two parts. 
% \shiqi{delete "s"?}

\begin{figure*}[t]
    \centering  
    \includegraphics[width=1\textwidth]{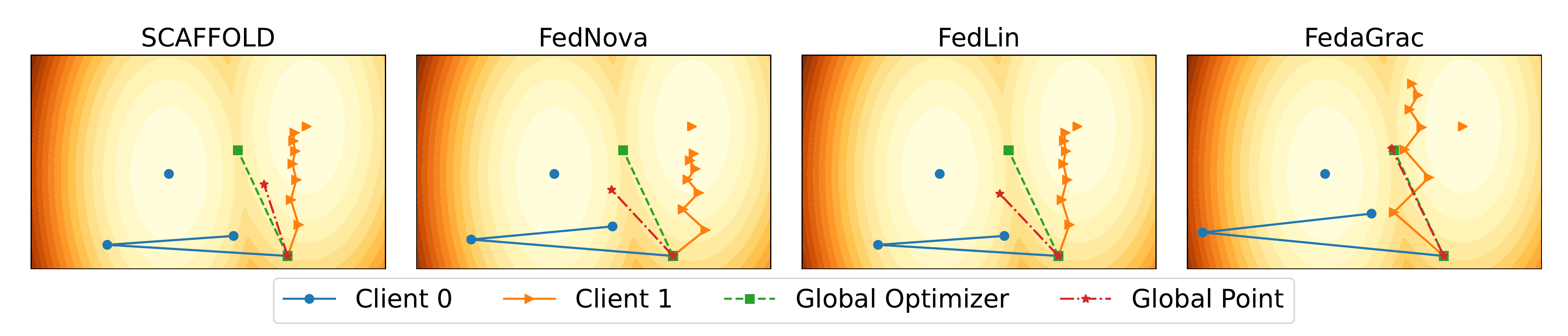}
    \caption{Illustration of model updating in the parameter space. For client $i$, we generate a set of $(x, y)$s fluctuated around a linear function $y = a_i x + b_i$, where $a_i$ and $b_i$ are real numbers. Our target is to find an optimal straight line $y = ax + b$, which is averagely close to all clients' data. Starting at the same point, each client applies \revise{mean squared error (MSE)} loss and follows a predefined algorithm to update its local model so as to optimize the global one. Regarding the existence of data and computation heterogeneity, our proposed method does not deviate from the direction towards a global minimizer.}
    \label{fig:pretest}
\end{figure*}

\noindent \textbf{Data Heterogeneity.} Generally, in FL settings, the data distributed among clients are agnostic and therefore, each data portfolio has its exclusive optimal parameters. As a classical algorithm to combat data heterogeneity, FedAvg inherits the training features from local SGD \cite{stich2018local, yu2019parallel, kavg}, a framework that runs for multiple local updates prior to a global synchronization. Obviously, this strategy significantly reduces the total communication overhead when compared to parallel SGD that synchronizes the gradient at every local update. Recent studies \cite{li2019convergence, khaled2020tighter, gu2021fast} show that FedAvg can have a great performance from theoretical and empirical perspectives. Also, FedAvg can seamlessly adopt communication-efficient approaches such as quantization \cite{alistarh2017qsgd, basu2019qsparse,wu2022sign} and sparsification \cite{stich2018sparsified, wangni2018gradient} to further reduce the cost of transmission \cite{zhou2021communication, wu2020fedscr, wang2021error, wu2020convergence}. 

Nevertheless, numerous studies \cite{karimireddy2020scaffold, zhao2018federated, liu2020accelerating, gorbunov2020unified, cheng2021dataset} theoretically prove that the issue raises the client-drift effect and degrades the convergence property. To mitigate the negative impact, existing solutions include cross-client variance reduction \cite{karimireddy2020scaffold, liang2019variance}, client clustering sampling \cite{ghosh2020efficient, murata2021bias, fraboni2021clustered} and reinforcement learning driven incentive mechanism \cite{wang2020optimizing}. Among these approaches, SCAFFOLD \cite{karimireddy2020scaffold} is a superior option that adjusts every local update with the help of the global and a client's local reference orientation, such that every local update keeps close to the global direction. 
% possesses the best performance in terms of test accuracy under homogeneous environments. 
However, as shown in Figure \ref{fig:pretest}, SCAFFOLD cannot completely remove the drift. A physical explanation for the result is that the local reference directions of the faster nodes with more number of local updates lead to a significant deviation from the orientation towards the local optimizer. Since the global reference direction is aggregated by clients' local ones, it is intuitively dominated by the faster nodes (see Figure \ref{fig:pretest}), which betrays its origin intention. Although we use a similar design philosophy that ensures every local update along with the global orientation, the global orientation consists of the gradient that depends on the number of local updates, either the normalized gradient or the initial gradient. 
% \shiqi{Not although here.}

\noindent\textbf{Computational Heterogeneity.} The computation capabilities vary among clients because they use different devices. To minimize the computation differences, some existing works adopt a client sampling strategy \cite{huang2020efficiency, deng2021auction, zhou2020petrel, wu2020accelerating}, where only a small portion of clients transmit the gradients to the server. Compared to the case that requires full-worker participation, this scheme reduces the total training time. However, there still exists resource underutilization as the fastest client should wait for others' completion. 

A practical solution is to adopt step asynchronism, where each client performs an inconsistent number of local updates. Although FedAvg with step asynchronism can converge to a stable point under non-convex objectives \cite{yu2019parallel}, Wang et al. \cite{wang2020tackling} point out that objective inconsistency takes place under quadratic function. To alleviate the challenge of computational heterogeneity, effective approaches are constituted with normalization-based approach FedNova \cite{wang2020tackling} and FedLin \cite{mitra2021linear}, regularization-based approach FedProx \cite{li2020federated} and architecture-based approach HeteroFL \cite{diao2020heterofl}. Gradient normalization is the most ubiquitous framework that overcomes step asynchronism under non-i.i.d. data setting. However, this method cannot prevent the negative impact of statistical heterogeneity on the convergence rate because the update deviation still exists after averaging. Figure \ref{fig:pretest} compares FedNova \cite{wang2020tackling} and FedLin \cite{mitra2021linear} with our proposed method, and we notice that the global model deviates to the one with less updates in FedNova \cite{wang2020tackling}. The reason is obvious: clients update the models bias to their local datasets such that the normalized gradients collected by the server are sparse. Besides, with the local models approaching the local minimizers, the update becomes so trivial that those clients with more local updates have a dispensable influence on the global model update. 

\section{Preliminary and Problem Formulation} \label{sec:preliminary}

Formally, the learning problem can be represented as the following distributed optimization problem across $M$ FL clients:
\begin{equation} \label{eq:problem}
    \min_{\mathbf{x}\in\mathbb{R}^d} F(\mathbf{x}) = \sum^M_{i = 1} \omega_i F_i(\mathbf{x}),
\end{equation}
\noindent where the weight $\omega_i = |\mathcal{D}_i|/|\mathcal{D}|$ is the ratio between the size of local dataset $\mathcal{D}_i$ and overall dataset $\mathcal{D}\triangleq\cup_{i=1}^M\mathcal{D}_i$, and $F_i(\mathbf{x})\triangleq\mathbb{E}_{\varepsilon_i\sim\mathcal{D}_i}[f_i(\mathbf{x};\varepsilon_i)]$ is the
the local objective, i.e., the expected loss value of model $x$ with respect to random sampling $\varepsilon_i$ for client $i$.

% \shiqi{FedAvg is not just aggregation? Also, naive is not appropriate here}
\textbf{FedAvg with step asynchronism.} Naive weighted aggregation \cite{mcmahan2017communication, stich2018local, li2019convergence, yu2019parallel} is an effective and communication-efficient way to solve Problem (\ref{eq:problem}) for both convex and non-convex objectives. With the increasing number of edge devices participating in model training, the framework is neither economic nor fair to require all clients to run a certain number of local updates. Instead, a practical approach is that client $i \in \{1, ..., M\}$ runs for a flexible number of SGD steps (i.e., $K_i$) according to its resource capability before the model aggregation at the server: 
% Server receives and averages the model parameters from clients, while active nodes proceed with the following steps in parallel: 
\begin{itemize}
    \item \textbf{(Pull):} Pulls the current parameter $\mathbf{x}_0$ from the server. 
    \item \textbf{(Compute):} Samples a realization $\varepsilon$ randomly from the local dataset $\mathcal{D}_i$ and compute the gradient $\nabla f_i(\mathbf{x}_k, \varepsilon)$. 
    \item \textbf{(Update):} Performs $k$-th local update of the form $\eta$ by $\mathbf{x}_{k+1} = \mathbf{x}_{k} - \eta \mathbf{g}_k$, where $k \in \{0, ..., K_i\}$ and $\eta$ is the stepsize.  
    \item \textbf{(Push):} Pushes the local parameter $\mathbf{x}_{K_i}$ to the server. 
\end{itemize}

Under this framework, we let $K_{\max}$ and $K_{\min}$ separately be the maximum and the minimum number of local updates among all clients, i.e., $K_{\max} = \max_{i \in \{1, ..., M\}} K_i$ and $K_{\min} = \min_{i \in \{1, ..., M\}} K_i$. In addition, $\bar{K} = \sum_{i=1}^M \omega_i K_i$ is defined as the weighted averaged number of local updates. Formally, step asynchronism is defined as the following mathematical expression: 
\begin{equation}
    \exists i, j \in \{1, ..., M\}, \quad K_i \neq K_j. 
\end{equation}
Therefore, $K_{\max} \neq K_{\min}$ when step asynchronism exists. Without extra explanations, these notations are adopted throughout the paper. 

\textbf{Assumptions.} To establish the convergence theory of the FL optimization, we make the following assumptions that are adapted in previous works \cite{li2019convergence, wang2020tackling, reddi2020adaptive, li2020federated, wang2019osp}: 

\begin{assumption}[L-smooth] \label{ass:1}
The local objective functions are Lipschitz smooth: For all $v, \Bar{v} \in \mathbb{R}^d$, 
\begin{equation*}
    \|\nabla F_i(v) - \nabla F_i(\Bar{v})\|_2 \le L\|v-\Bar{v}\|_2, \quad \forall i \in \{1, ..., M\}.
\end{equation*}
\end{assumption}

\begin{assumption}[$\mu$-strongly convex] \label{ass:2}
The local objective functions are $\mu$-strongly convex with the value of $\mu > 0$: For all $v, \Bar{v} \in \mathbb{R}^d$, 
\begin{equation*}
\begin{split}
    F_i(v) - F_i(\Bar{v}) \geq \langle \nabla F(\Bar{v}), (v-\Bar{v}) \rangle + \frac{\mu}{2} \|v-\Bar{v}\|_2^2, \quad \\
    \forall i \in \{1, ..., M\}
\end{split}
\end{equation*}
where $\langle \cdot, \cdot \rangle$ refers to the inner product of two gradients. 
\end{assumption}

\begin{assumption}[Bounded Variance] \label{ass:3}
For all $v \in \mathbb{R}^d$, there exists a scalar $\sigma \geq 0$ such that 
\begin{equation*}
    \mathbb{E}\|\nabla f_i(v,\varepsilon) - \nabla F_i(v)\|_2^2 \le \sigma^2, \quad \forall i \in \{1, ..., M\}
\end{equation*}
\end{assumption}

\begin{assumption}[Bounded Dissimilarity] \label{ass:4}
For some $v \in \mathbb{R}^d$ that $\|\nabla F(v)\|_2^2 > 0$ holds, there exists a scalar $B \geq 1$ such that 
\begin{equation*}
    \mathbb{E}\|\nabla F_i(v)\|_2^2 \le B^2 \|\nabla F(v)\|_2^2, \quad \forall i \in \{1, ..., M\}.
\end{equation*}
Obviously, when the data are independent and identically distributed, the value of $B$ should be 1. 
\end{assumption}

Assumption \ref{ass:4} seems to be a little bit strong as $\|\nabla F(v)\|_2^2$ cannot be 0. However, considering $\epsilon$-accuracy as the learning criterion, i.e., $\|\nabla F(v)\|_2^2 \leq \epsilon$ under non-convex objectives \revise{such as deep neural networks which possess multiple local minimizers}, the value of $\epsilon$ cannot strictly be 0. In other words, there exists $\epsilon_1 \leq \epsilon$ such that $\|\nabla F(v)\|_2^2 \geq \epsilon_1$ for all $v$ always holds. 

\textbf{Key factor that raises objective inconsistency.} Although \cite{wang2020tackling} indicates that objective inconsistency occurs when using FedAvg with step asynchronism under quadratic functions, the factor that makes it happen remains a mystery. To explore in depth, the following theorem analyzes FedAvg with step asynchronism under a strongly-convex objective. 

\begin{theorem}\label{tm:tm1}
Suppose the local objective functions are non-negative. Denote the parameter at $t$-th communication round by $\mathbf{x}_t$. Let $T$ be the total number of communication rounds. Under Assumption \ref{ass:1}, \ref{ass:2}, \ref{ass:3} and \ref{ass:4}, by setting the learning rate $\eta = \mathcal{O}(1/\mu LT\bar{K}) \leq 1/L\bar{K}$, the output of FedAvg with step asynchronism satisfies 
% \begin{equation} \label{eq:fedavg_with_step_asyn}
% \begin{split}
%     &\quad\mathbb{E}[F(\mathbf{x}_T)] - F(\mathbf{x}_*) \\
%     &\leq \tilde{\mathcal{O}}\left(\frac{\mu \bar{K}}{K_{\min}} \|\mathbf{x}_1 - \mathbf{x}_*\|_2^2 \exp\left({-\frac{\mu T}{L}}\right)\right. \\
%     &\qquad + \frac{\sigma^2}{\mu T \bar{K} K_{\min}} \sum_{i=1}^M \omega_i^2 K_i + \frac{\sigma^2 L}{\mu^2 T^2 \bar{K}^2 K_{\min}} \sum_{i=1}^M \omega_i K_i^2\\
%     & \qquad \left. + \sum_{i=1}^M \omega_i \left(\frac{K_i}{K_{\min}} - 1\right) F_i(\mathbf{x}_*) \right).
% \end{split}
% \end{equation}
\begin{equation}\label{eq:fedavg_with_step_asyn}
    \lim_{T \rightarrow \infty} \mathbb{E}[F(\mathbf{x}_T)] - F(\mathbf{x}_*) \leq O\left(\sum_{i=1}^M \omega_i \left(\frac{K_i}{K_{\min}} - 1\right) F_i(\mathbf{x}_*)\right)
\end{equation}
where $\mathbf{x}_1$ and $\mathbf{x}_*$ indicates the initial and optimal model parameters, respectively. 
\begin{proof}
See Appendix \ref{proof:tm1} for details. 
% See Appendix A for details. 
\end{proof}
\end{theorem}
\noindent\textbf{Remark} \quad The theoretical result in Equation \ref{eq:fedavg_with_step_asyn} is consistent with the result of FedAvg analysis in \cite{karimireddy2020scaffold} as the number of local updates is identical, i.e., $K_{i} = \bar{K}, \forall i \in \{1, ..., M\}$. Besides, when the data are identical and independent distributed among clients, where the global optimizer is not equivalent to the clients' local minimizer, we can easily induce that $\mathbf{x}_T$ is close to $\mathbf{x}_*$ when $T \rightarrow \infty$. The conclusion holds regardless of the number of local updates. However, when data heterogeneity and step asynchronism coexist, the right-hand side of Equation (\ref{eq:fedavg_with_step_asyn}) is non-zero. As a result, when $T$ tends to be infinite, the model cannot converge to the optimal parameters, which can explain the result manifested in Table \ref{table:fedavg} under LR. Based on the theoretical discovery, we can draw a conclusion that step asynchronism leads to a significant accuracy drop in the non-i.i.d. cases, which impedes normal training. 

\section{\texttt{FedaGrac} algorithm} \label{sec:FedaGrac}

\noindent To ensure that $\mathbb{E}[F(\mathbf{x}_T)] - F(\mathbf{x}_*)$ is close to 0 when $T \rightarrow \infty$, we target to remove the constant term in the right-hand side of Equation (\ref{eq:fedavg_with_step_asyn}). Based on the remark in Section \ref{sec:preliminary}, a practical approach is to minimize the effectiveness of data heterogeneity. 
% As mentioned in the related work, we intend to eliminate the negative effect raised from data heterogeneity such that each local update does not critically deviate from the expected orientation.
In this section, we elaborate our proposed algorithm, \underline{Fed}erated \underline{A}ccelerating \underline{Gra}dient \underline{C}alibration (\texttt{FedaGrac}), to avoid the objective inconsistency as well as enhance the convergence performance when step asynchronism is adopted to improve the resource utilization. The implementation details are presented as Algorithm \ref{algo:1}. 
% -- from two different dimensions, namely, calibrating the local client deviation and estimating the global reference orientation

At first, apart from the hyperparameters such as learning rate $\eta$ and calibration rate $\lambda$, we initialize a $d$-dimension model with arbitrary parameters $\mathbf{x}_1$. Besides, to ease the theoretical analysis in Section \ref{sec:theory}, we set $\nu^{(i)}$ as $\nabla f_i(\mathbf{x}_1, \mathcal{D}_i)$ for all $i \in \{1, ..., M\}$. Then, we define $\nu$ as: 
% Then, $\nu$ begins with the following formula \shiqi{Then, we define $\nu$ as}:
\begin{equation*}
    \nu = \sum_{i=1}^M \omega_{i} \nu^{(i)} = \sum_{i=1}^M \omega_{i} \nabla f_i(\mathbf{x}_1, \mathcal{D}_i). 
\end{equation*}
In this algorithm, client $i$ performs the local updates for $K_i$ times in parallel. During each local update, clients calibrate the local client deviation with reference to the global reference orientation, which is estimated at every global synchronization. In the following two subsections, we separately discuss the effectiveness of two main components, namely, 
\begin{itemize}
    \item \textbf{Calibrating the local client deviation} (Line 9 in Algorithm \ref{algo:1}) migrates the data heterogeneity; 
    \item \textbf{Estimating the global reference orientation} (Line 14 in Algorithm \ref{algo:1}) accelerates the training process. 
\end{itemize}
% calibrating the local client deviation for migrating the data heterogeneity, and estimating the global reference orientation for accelerating the training process. 

\begin{algorithm}[!t]
  \caption{\underline{Fed}erated \underline{A}ccelerating \underline{Gra}dient \underline{C}alibration (\texttt{FedaGrac})}
  \label{algo:1}
  \begin{algorithmic}[1]
    \REQUIRE
        Initialize $M$ clients, set the initial model to be $\tilde{\mathbf{x}}_1\in\mathbb{R}^d$. Set $\nu^{(i)}$ and $\nu$ for all clients $i \in \{1, \cdots, M\}$. Set learning rate $\eta > 0$, calibration rate $\lambda > 0$, the number of global synchronizations $T$ and the number of local iterations of each client $K_i$ for all clients $i \in \{1, \cdots, M\}$.
    \STATE \underline{{\bfseries On client $i\in\{1,\cdots,M\}$}:}
    % \STATE Initialize $\nu^{(i)} = \mathbf{0}$. 
    \FOR{$t = 1$ to $T$}
    {
        \STATE Pull $\tilde{\mathbf{x}}_t, \nu$ from server 
        \STATE Set $\mathbf{x}^{(i)}_{t;0} = \tilde{\mathbf{x}}_t$
        % \STATE Set $\mathbf{c} = \nu - \nu^{(i)}$
        \STATE Set $\mathbf{c} = \nu - \nu^{(i)}$
        \FOR{$k = 0$ to $K_i - 1$}
        {
            \STATE Randomly sample a realization $\varepsilon^{(i)}_k$ from  $\mathcal{D}_i$
            \STATE $\mathbf{g}^{(i)}_{t;k} = \nabla f_i(\mathbf{x}^{(i)}_{t;k}, \varepsilon^{(i)}_k)$
            \STATE $\mathbf{x}^{(i)}_{t;k+1} = \mathbf{x}^{(i)}_{t;k} - \eta (\mathbf{g}^{(i)}_{t;k} + \lambda \mathbf{c})$ 
            % \STATE $\mathbf{x}^{(i)}_{t;k} = \mathbf{x}^{(i)}_{t;k-1} - \eta_k \mathbf{g}^{(i)}_{t;k}$
        }
        \ENDFOR
        \STATE Set $\nu^{(i)} = \frac{1}{K_i} \sum_{k=0}^{K_i-1} g_{t;k}^{(i)}$
        \STATE Push $\mathbf{x}^{(i)}_{t;K_i}, K_i$ to the server
        \STATE Receive $\bar{K}$ from the server
        \STATE if $K_i \leq \bar{K}$, then send $\nu^{(i)}$; else send $g_{t;0}^{(i)}$
    }
    \ENDFOR   
    \STATE \underline{{\bfseries On server}:}
    % \STATE Initialize $\nu = \mathbf{0}$. 
    \FOR{$t = 1$ to $T$}
    {
        \STATE Push $\tilde{\mathbf{x}}_t, \nu$ to clients 
        \STATE Pull $\mathbf{x}^{(i)}_{t;K_i}, K_i$ from client $i\in[1,\dots,M]$
        \STATE $\tilde{\mathbf{x}}_{t+1}$ = $\sum_{i=1}^M \omega_i \mathbf{x}^{(i)}_{t;K_i}$
        \STATE $\bar{K} = \sum_{i=1}^M \omega_i K_i$
        \STATE Push $\bar{K}$ to clients and receive $v^{(i)}_{\text{transit}}$ from clients 
        \STATE $\nu = \sum_{i=1}^M \omega_i v^{(i)}_{\text{transit}}$
    }
    \ENDFOR
  \end{algorithmic}
\end{algorithm}

\subsection{Calibrating the local client deviation}

\begin{figure}[t]
    \vspace{8px}
    \captionsetup[subfloat]{farskip=2pt,captionskip=1pt}
    \centering
    \subfloat[w/ step async]{
        \raisebox{-.5\height}{\includegraphics[width=.3\textwidth]{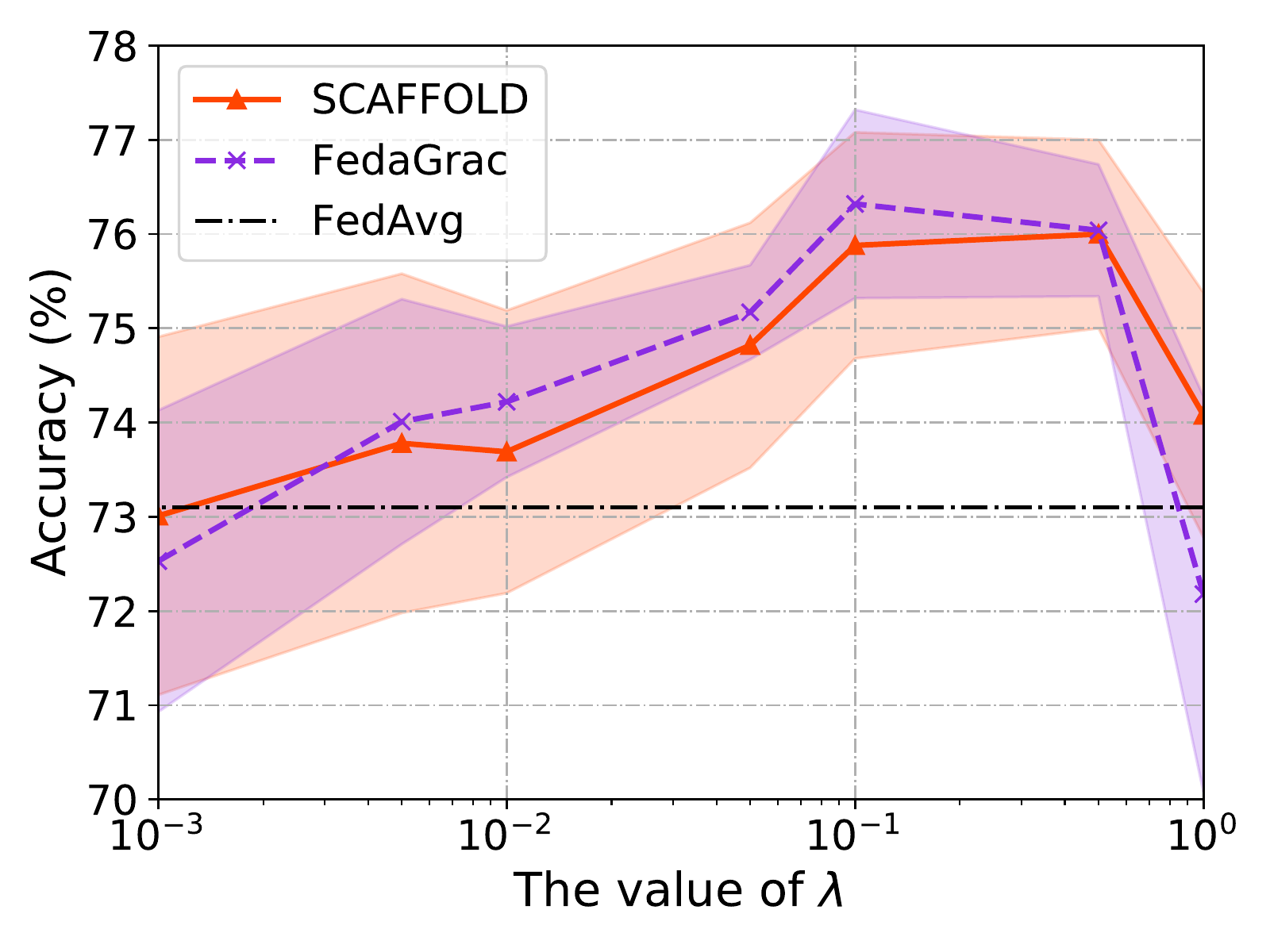}}
        \label{fig:lam_step}
    }
    \subfloat[w/o step async]{
        \vspace*{10px}
        \begin{tabular}{cc}
        \hline
            $\lambda$ & Acc. (\%) \\ \hline
            0 & 77.37$\pm$0.8 \\
            0.001 & 77.23$\pm$1.2 \\
            0.005 & 77.18$\pm$0.9 \\
            0.01 & 77.49$\pm$0.6 \\
            0.05 & {78.23$\pm$0.5} \\
            0.1 & {78.26$\pm$0.3} \\
            0.5 & 77.03$\pm$0.7 \\
            1 & 77.29$\pm$0.9 \\ 
            \footnotesize{Increase} & \textbf{79.21$\pm$0.4} \\ \hline
            \vspace{-6px}
        \end{tabular}
        \label{tab:lam_step}
    }
    % \vspace{-10px}
    \caption{Preliminary experiments for a 2-layer CNN with the recognition of Fashion-MNIST. In the line graph \ref{fig:lam_step}, the experiments are conducted when the clients perform various numbers of local steps. The vertical axis indicates the test accuracy after 200 rounds, while the horizontal axis shows the value of calibration rate $\lambda$. As for the table \ref{tab:lam_step}, clients perform constant local updates. The algorithm is equivalent to FedAvg when $\lambda=0$. ``Increase'' in the table shows the value of $\lambda$ changes over time, i.e., 0.1 for the first 50 rounds, 0.5 for the next 100 rounds, and 1 for the rest.}
    % , where SCAFFOLD and FedaGrac are the same
    \label{fig:various_lam}
    % \vspace{-5px}
\end{figure}

% A practical approach to calculate the gradients is based on the empirical results, where Client $i$ randomly samples a mini-batch from $\mathcal{D}_i$ and computes on the model $\x{i}{t}{k}$ using the loss function such as cross entropy loss. In this case, the empirical gradient should be $\mathbf{g}^{(i)}_{t;k}$, which is unbiased estimated to $\grad{i}{t}{k}$ in expectation. Obviously, following $\x{i}{t}{k+1} = \x{i}{t}{k} - \eta \grad{i}{t}{k}$ to update the local model undoubtedly deviates to the local optimizer, which seriously affects the convergence property. What is worse, Section \ref{sec:preliminary} concludes that objective inconsistency cannot be overcome unless data heterogeneity is solved. 

As a classical approach, FedAvg updates the parameters using stochastic gradient descent (SGD), where the gradient is computed in accordance with Line 8. Suppose the gradient is equivalent to the first order derivative of the true local objective, i.e., $\nabla f_i(v, \varepsilon) = \nabla F_i(v)$, where $\varepsilon$ is randomly sampled from the local dataset $\mathcal{D}_i$. Then, given the stepsize $\eta$, the model update follows $\x{i}{t}{k+1} = \x{i}{t}{k} - \eta \grad{i}{t}{k}$. Previous works \cite{zinkevich2010parallelized, bottou2010large} show that this scheme can converge to a stable point when a client performs sufficient local updates. Under a heterogeneous data setting, the points vary among clients because each of them is determined by the local data distribution. Therefore, clients are biased from the global orientation, and the phenomenon is named as client deviation\footnote{Client deviation is also known as client drift \cite{karimireddy2020scaffold, mitra2021linear}.}. 

% Client deviation, also known as client-drift \cite{karimireddy2020scaffold}, refers to an event that a client deviates from the expected direction during its local updates. 
Existing works to overcome client deviation mainly focus on the variance reduction approach, i.e., SCAFFOLD \cite{karimireddy2020scaffold}. It is somehow similar to our proposed algorithm when $\lambda$ is set to 1. In this case, client $i$'s local reference direction $v^{(i)}$ is assumed to be equivalent to the vector from the current point to its local optimizer. As for the global reference orientation, $v$ overlaps with the gradient from the current point to the global minimizer. However, it is nearly impossible to coincide with the case, especially when applied with the gradient calibration technique. Generally speaking, using an obsolete gradient to predict the coming gradient is not reasonable because the aggregated direction presumably deviates from the expected one. 

% Different from SCAFFOLD, we define a calibration rate $\lambda$ for the correction term. In an ideal case that $\lambda = 1$, Client $i$'s local reference direction $v^{(i)}$ is equivalent to the vector from current point to its local optimizer, while the global reference orientation $v$ overlaps with the vector from current point to the global minimizer. However, it is nearly impossible to coincide with the case, especially when applied with gradient calibration technique. Generally speaking, using an obsolete gradient to predict the coming gradient is not reasonable because the aggregated direction presumably deviates from the expected one. As a result, we utilize a hyper-parameter to adjust the calibration technique as well as enhance the feasibility. 

% Suppose all clients involve in the training all the time, for $t$-th round, $\nu^{(i)}$ has been estimated based on the model $\x{t}$ (Line 14 and Line 16), and therefore $\nu_t$ represents the approximated global orientation for $t$-th round. Considering the case that solely a fraction of clients (denote by $\mathcal{A}$) involve in the training, we migrate the caching orientations for clients who do not participate in $(t-1)$-th round training from the results they uploaded last time. Therefore, we devise Line 18. For each local update, client $i$ follows the rule $\x{i}{t}{k+1} = \x{i}{t}{k} - \eta \bracket{\mathbf{g}^{(i)}_{t;k} - \nu^{(i)} + \nu_t} \overset{(a)}{\approx} \x{i}{t}{k} - \eta \nabla F(x_{t, k}^{(i)})$, where $(a)$ holds as long as there is no noise and all workers participating in the training. 

Therefore, we introduce a calibration rate $\lambda$ for the correction term. With this hyperparameter, a gradient can be adjusted and approximated to the global update. Empirical results in Figure \ref{fig:various_lam} intuitively present the effectiveness of $\lambda$. Generally speaking, a smaller $\lambda$ has a similar performance as FedAvg because the calibrated gradient is still biased to the local computed one. For a greater $\lambda$, the test accuracy goes down dramatically since the gradient is over-calibrated. As a result, a constant $\lambda$ cannot be too large or too small such that the calibration term is effective. Furthermore, in Figure \ref{tab:lam_step}, we evaluate a case where $\lambda$ increases over time. Apparently, the strategy is impressive because it outperforms all other constant settings. The reason for the improvement is clear: at the beginning stage, the difference between two successive updates is significant because the model is far away from convergence. When the training comes to a stable point, the value of $\lambda$ should be 1 such that the gradient eliminates the deviation towards the local minimizer. 

\subsection{Estimating the global reference orientation}

\begin{figure}[t]
\centering
\subfloat[LR w/o step async] {\includegraphics[width=.23\textwidth]{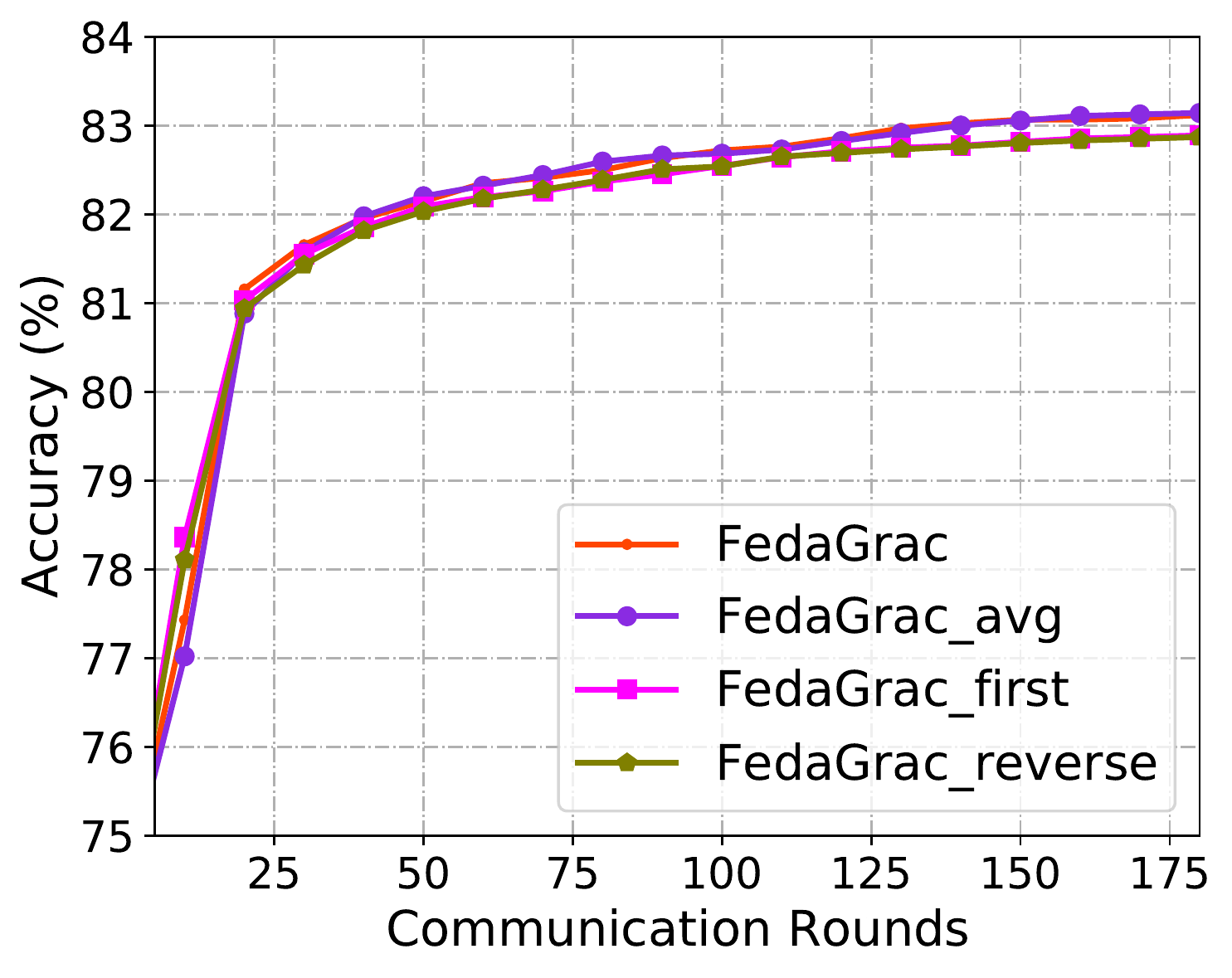} \label{fig:fedagrac_lr_nonstep}}
\subfloat[2-layer CNN w/o step async] {\includegraphics[width=.23\textwidth]{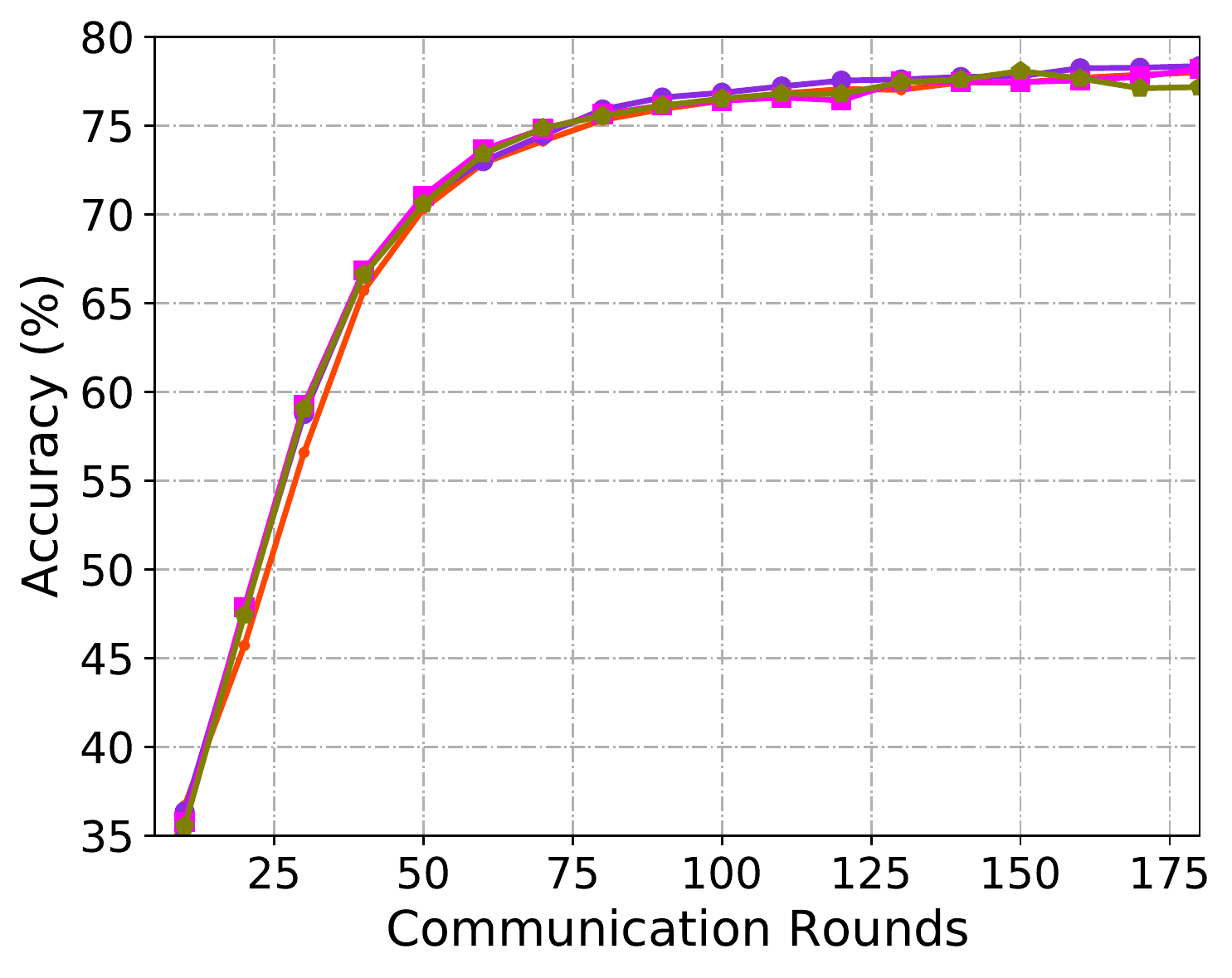} \label{fig:fedagrac_cnn_nonstep}}
\\
\subfloat[LR w/ step async] {\includegraphics[width=.23\textwidth]{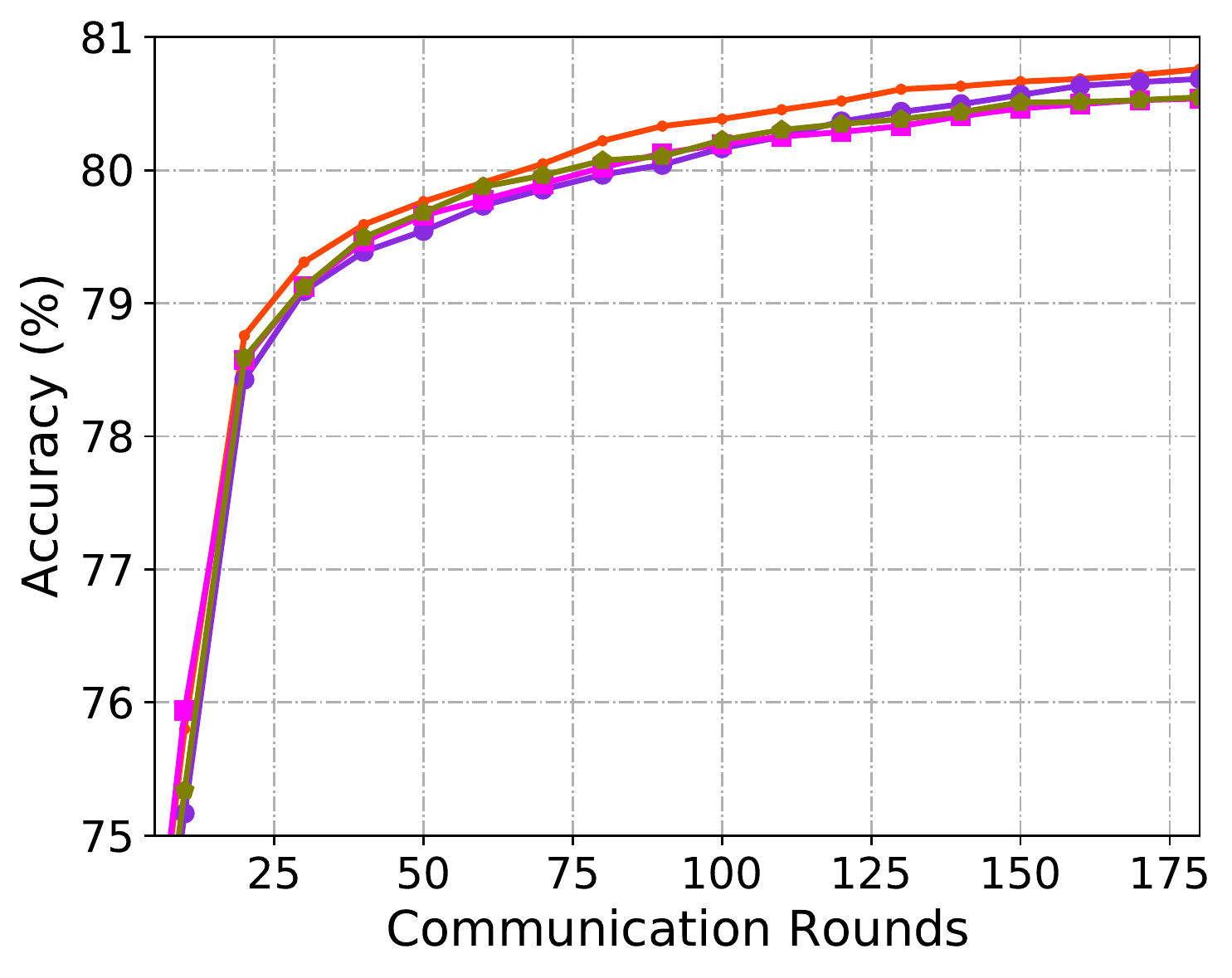}\label{fig:fedagrac_lr_step}}
\subfloat[2-layer CNN w/ step async] {\includegraphics[width=.23\textwidth]{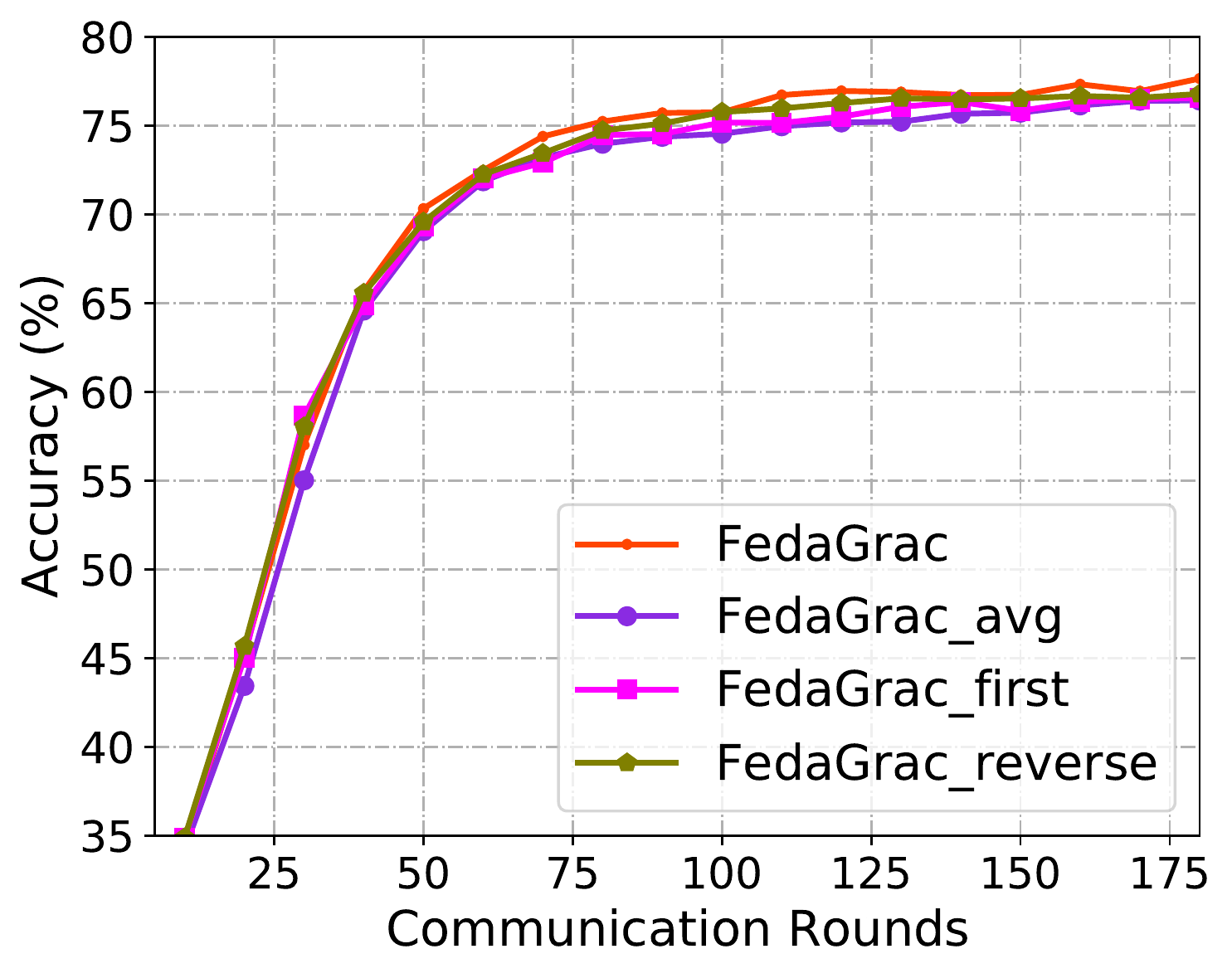}\label{fig:fedagrac_cnn_step}}
\caption[t]{Empirical evaluation for how to estimate the global reference orientation using Fashion-MNIST with convex (i.e., LR) and non-convex objectives (i.e., 2-layer CNN). The horizontal axis indicates the communication rounds, and the vertical axis shows the test accuracy in percentage. (a)(b) indicate the results when the clients run for the constant number of updates, and (c)(d) is when they perform various numbers of SGD steps. (Zoom in for the best view)}
% \shiqi{Scale y-axis. Cannot distinguish lines.}
\label{fig:fedagrac_evaluation}
\end{figure}

While applying SCAFFOLD \cite{karimireddy2020scaffold} to train a model, we notice that the model update is biased to the fastest node under step-asynchronous settings. Given a model $\mathbf{x}$, some clients, e.g., client $i$, are close to a stable point such that the computed local reference orientation significantly deviates from the expected one, i.e., $\nabla F_i(\mathbf{x})$. Regarding that the clients (client $i$) with fewer local updates can better estimate the local orientation $\nabla F_i(\mathbf{x})$, the model prefers those with more local updates, which undermines the convergence property. 

At the beginning of round $t \in \{1, ..., T\}$, the centralized server broadcasts the model $\tilde{\mathbf{x}}_t$ to all clients. To obtain an exact result of $\nabla F(\tilde{\mathbf{x}}_t)$, each client $i \in \{1, ..., M\}$ should provide an accurate estimation for $\nabla F_i(\tilde{\mathbf{x}}_t)$, or the bias of the estimation $\nu^{(i)}$ can be eliminated by the sum, i.e., $\sum_{i=1}^M \omega_i \nu^{(i)}$. Therefore, there are two practical ways to estimate $\nabla F_i(\tilde{\mathbf{x}}_t)$ for client $i$, namely, (i) the first stochastic gradient, i.e., $\nabla f_i(\tilde{\mathbf{x}}_t, \varepsilon)$, and (ii) the averaged stochastic gradient, i.e., $ \frac{1}{K_i} \sum_{k=0}^{K_i-1} \nabla f_i\left(\mathbf{x}_{t,k}^{(i)}, \varepsilon_k^{(i)}\right)$ in Line 11 of Algorithm \ref{algo:1}. Based on these two strategies, we design and empirically evaluate four different schemes to find a proper estimation for the global reference orientation: (Note: faster or slower nodes are classified by whether the number of local updates is greater than the average updates)
\begin{itemize}
    \item \textbf{FedaGrac} requires faster nodes to transmit the first stochastic gradient while the rest push the average one; 
    \item \textbf{FedaGrac\_avg} (a.k.a. SCAFFOLD) requires all nodes to transmit the average stochastic gradient; 
    \item \textbf{FedaGrac\_first} requires all nodes to transmit the first stochastic gradient;
    \item \textbf{FedaGrac\_reverse} requires faster nodes to transmit the average stochastic gradient while the rest push the first one. 
\end{itemize}

Figure \ref{fig:fedagrac_evaluation} presents the results of different strategies. As we can see, without step asynchronism, these four schemes do not have considerable differences. However, with step asynchronism, \texttt{FedaGrac} outperforms another three potential approaches under both convex and non-convex objectives. This is why Line 14 of Algorithm \ref{algo:1} is introduced. 
To further reduce the communication overhead, the algorithm solely requests the faster nodes to upload the first stochastic gradient, while the rest can be computed via $\frac{1}{\eta K_i} \left(\tilde{\mathbf{x}}_t - \mathbf{x}_{t, K_i}^{(i)}\right) - \lambda \left(\nu - \nu^{(i)}\right)$ if $\nu^{(i)}$ is preserved on the server. 

\section{Theoretical Convergence Analysis} \label{sec:theory}

In this section, we analyze the convergence property of \texttt{FedaGrac} under both non-convex objectives and strongly-convex objectives for solving Problem (\ref{eq:problem}). The details of the mathematical proof are provided in the supplementary materials with step-by-step explanations. 

\subsection{Mathematical expression for Algorithm \ref{algo:1}}

In Section \ref{sec:FedaGrac}, we describe the details in Algorithm \ref{algo:1}. Below represents how to derive the recursive function step by step. 

\textbf{Local reference orientation.} To ensure every local update can calibrate to the expected one, we should use the averaged local update such that after multiple local updates, the acquired model does not deviate from the expected orientation. Therefore, the local reference orientation is defined as: 
\begin{equation}
    \nu^{(i)}=\begin{cases}
    \frac{1}{K_i} \sum_{k=0}^{K_i-1} g_{t-1;k}^{(i)}, & K_i \leq \bar{K}\\
    g_{t-1;0}^{(i)}, &\text{Otherwise}
    \end{cases}
\end{equation}

\textbf{Global reference orientation.} SCAFFOLD \cite{karimireddy2020scaffold} presents a remarkable performance with the aggregation of $\nu^{(i)}$ for all $i \in \{1, ..., M\}$. However, the approach presumably does not work due to step asynchronism, where local reference orientations deviated from the expected direction are dramatically various among clients. To avoid this issue, we let the faster node with more number of local updates transfer the initial gradient while others send the local reference orientation to the server, which can be formally written as: 
\begin{align*}
    \nu &=\sum_{i, K_{i} \leq \bar{K}} \frac{\omega_i}{K_{i}} \sum_{k=0}^{K_{i}-1} g_{t-1 ; k}^{(i)}+\sum_{i, K_{i}>\bar{K}} \omega_i g_{t-1; 0}^{(i)}
    % &=\sum_{i=1}^M \frac{\w_i}{K_{i}} \sum_{k=0}^{K_{i}-1} g_{t-1;k}^{(i)} + \sum_{i, K_{i} > \bar{K}} \frac{\w_i}{K_{i}} \sum_{k=0}^{K_i-1}\bracket{g_{t-1; 0}^{(i)}-g_{t-1; k}^{(i)}} \\
    % &=\sum_{i, K_{i} \leq \bar{K}} \frac{\w_i}{K_{i}} \sum_{k=0}^{K_{i}-1}\bracket{g_{t-1;k}^{(i)}-g_{t-1;0}^{(i)}}+\sum_{i=1}^{M} \w_i g_{t-1; 0}^{(i)}
\end{align*}

\textbf{Recursion function.} According to Line 9 in Algorithm \ref{algo:1}, for client $i$, the recursion between two successive local updates can be presented as: 
\begin{align} \label{eq:pre-1}
    \x{i}{t}{k+1} = \x{i}{t}{k} - \eta \left[g_{t; k}^{(i)}+\lambda \bracket{\nu-\nu^{(i)}}\right]
\end{align}
Then, based on the equation above, i.e., Equation (\ref{eq:pre-1}), for client $i$ with the local updates of $K_i$, $\x{i}{t}{K_i} - \gx{t}$ can be formulated in mathematical expression as: 
\begin{align*}
    \x{i}{t}{K_i} - \gx{t} &= \sum_{k=0}^{K_i -1} \bracket{\x{i}{t}{k+1}-\x{i}{t}{k}}\\
    &=- \eta \sum_{k=0}^{K_i-1} g_{t;k}^{(i)}-\eta \lambda K_i \bracket{\nu-\nu^{(i)}}
\end{align*}
Finally, according to the definition in Problem (\ref{eq:problem}), the recursion function between two successive global updates is the weighted average of all clients' models, which is written as: 
\begin{align*}
    \gx{t+1} - \gx{t} &= \sum_{i=1}^M \w_i\x{i}{t}{K_i} - \gx{t}  \\
    &= - \eta \sum_{i=1}^M \sum_{k=0}^{K_i-1} \w_i g_{t;k}^{(i)}-\eta \lambda \bar{K} \nu + \eta \lambda \sum_{i=1}^M \w_i K_i \nu^{(i)}
\end{align*}

\subsection{Non-convex objectives}

\begin{theorem}[Non-convex objectives] \label{tm:tm2}
Considering the same $\mathbf{x}_1$ and $\mathbf{x}_*$ as Theorem~\ref{tm:tm1}, under Assumption \ref{ass:1}, \ref{ass:3} and \ref{ass:4}, by setting $\eta = \mathcal{O}\left(\sqrt{\frac{M}{T\bar{K}}}\right)$, the convergence rate of Algorithm~1 with step asynchronism for non-convex objectives is
%\begin{equation}
%\begin{split}
%    &\quad\frac{1}{T} \sum_{t=1}^T \mathbb{E} \|\nabla F(\tilde{\mathbf{x}}_t)\|_2^2\\ 
%    &\leq \mathcal{O}\left(\frac{F(\tilde{\mathbf{x}}_1) - F(\mathbf{x}_{*})}{\eta \lambda \bar{K} T}\right) + \mathcal{O}\left(\frac{\eta \sigma^2 L}{\lambda \bar{K}} \sum_{i=1}^M \omega_i^2 K_i\right) \\
%    &\quad+\mathcal{O}\left(\eta \sigma^2 L\lambda \sum_{i=1}^M \omega_i^2 \left(\frac{(\bar{K}-K_i)^2}{\bar{K}K_i} + 1\right) \right)\\
%    &\quad + \mathcal{O}\left(\frac{\eta^2 L^2 \sigma^2}{\lambda \bar{K}} \sum_{i=1}^M \omega_i K_i^4\right)\\
%    &\quad + \mathcal{O}\left(\frac{\eta^2 L^2 \sigma^2 \lambda}{\bar{K}} \sum_{i=1}^M \omega_i K_i^3 \left(K_i \sum_{j=1}^M \frac{\omega_j^2}{K_j} + 1\right)\right)
%\end{split}
%\end{equation}
\begin{align}
    	&\quad\frac{1}{T} \sum_{t=1}^T \mathbb{E} \|\nabla F(\tilde{\mathbf{x}}_t)\|_2^2 \nonumber\\ 
    	&\leq \mathcal{O}\left(\frac{(F(\mathbf{x}_1) - F(\mathbf{x}_{*}))}{\lambda \sqrt{\bar{K}MT}}\right) + \mathcal{O}\left(\frac{\sigma^2 L\sqrt{M}}{\lambda \sqrt{\bar{K}^{3}T}} \sum_{i=1}^M \omega_i^2 K_i\right) \nonumber\\
    	&\quad+\mathcal{O}\left(\frac{\sigma^2 L\lambda \sqrt{M}}{\sqrt{\bar{K}T}} \sum_{i=1}^M \omega_i^2 \left(\frac{(\bar{K}-K_i)^2}{\bar{K}K_i} + 1\right) \right) \nonumber\\
    	&\quad + \mathcal{O}\left(\frac{L^2 \sigma^2M}{\lambda \bar{K}^2T} \sum_{i=1}^M \omega_i K_i^4\right) \nonumber\\
    	&\quad + \mathcal{O}\left(\frac{L^2 \sigma^2 \lambda M}{\bar{K}^2T} \sum_{i=1}^M \omega_i K_i^3 \left(K_i \sum_{j=1}^M \frac{\omega_j^2}{K_j} + 1\right)\right).
	\end{align}
\begin{proof}
  % See Appendix \ref{proof:tm2} for details. 
See Appendix \ref{proof:tm2} for details. 
\end{proof}
\end{theorem}

\begin{corollary}
By setting $\omega_1 = ... = \omega_M = 1/M$ and $\lambda = \mathcal{O}(1)$, the following inequality holds under Theorem \ref{tm:tm2}: 
\begin{equation} \label{eq:non_convex_opt}
    \min_{t \in \{1, ..., T\}} \mathbb{E} \|\nabla F(\tilde{\mathbf{x}}_t)\|_2^2 \leq \mathcal{O} \left(\frac{1}{\sqrt{MT\bar{K}}}\right).
\end{equation}
\end{corollary}

\noindent\textbf{Remark}\quad \cite{wang2020tackling} states that FedaNova can achieve the convergence rate same as Equation \ref{eq:non_convex_opt}, but there exists an explicit condition that $\sum_{i=1}^M (\bar{K}/MK_i)$ is a constant when $\omega_1 = ... = \omega_M = 1/M$. Let us consider an extreme case that the slow nodes locally update once, i.e., $K_i = 1$ for all $i \in \{1, ..., M-1\}$ while Client $M$ can run for a very large number of times. This case is possible, for instance, a system consists of multiple Raspberry Pi and a single Nvidia GTX 3080Ti GPU, the computational difference between which can be up to a thousandfold. Under such situation, the aforementioned term should be bounded by $\mathcal{O}(\bar{K})$ instead of $\mathcal{O}(1)$ and therefore, the convergence rate for FedNova should be $\mathcal{O}(\sqrt{\bar{K}/MT})$. In comparison with Equation \ref{eq:non_convex_opt}, \texttt{FedaGrac} achieves an increment up to $\mathcal{O}(\bar{K})$. 

Furthermore, the algorithms such as FedAvg \cite{yu2019parallel} and SCAFFOLD \cite{karimireddy2020scaffold} that use the homogeneous setting achieve a convergence rate of $\mathcal{O}(1/\sqrt{MTK_{\min}})$. Obviously, \texttt{FedaGrac} admits better convergence rate as $K_{\min} \leq \bar{K}$ always holds under heterogeneous computational resources. This is because our algorithm can fully utilize the computational resources from all participants such that it outperforms those algorithms that solely supports the homogeneous environment.

\subsection{Strongly-convex objectives}

\begin{theorem}[Strongly-convex objectives]
\label{tm:tm3}
Considering the same $\mathbf{x}_1$ and $\mathbf{x}_*$ as Theorem~\ref{tm:tm1}, under Assumption \ref{ass:1}, \ref{ass:2} and \ref{ass:3}, by setting $\lambda = 1$, $\eta = \mathcal{O}(1/\mu LT\bar{K}) \leq 1/L\bar{K}$, the convergence rate of Algorithm~1 with step asynchronism for strongly-convex objectives is
\begin{equation}
\begin{split}
    &\quad\mathbb{E}[F(\tilde{\mathbf{x}}_T)] - F(\mathbf{x}_*) \\
    &\leq \tilde{\mathcal{O}}\left(\mu \|\mathbf{x}_1 - \mathbf{x}_*\|_2^2 \exp{\left(-\frac{\mu T}{L}\right)} + \frac{\mathcal{H}}{\mu T} + \frac{\mathcal{P}}{\mu^2 T^2}\right),
\end{split}
\end{equation}
where 
\begin{equation*}
\begin{split}
    \mathcal{H} &= \frac{\sigma^2}{\bar{K}^2} \sum_{i=1}^M \omega_i^2 \left(K_i + \bar{K} + \frac{(\bar{K} - K_i)^2}{K_i}\right),\\
    \mathcal{P} &= \frac{L^2 \sigma^2}{\mu} \left(\sum_{i=1}^M \omega_i K_i^3\right) \sum_{j=1}^M \frac{\omega_j^2}{K_j}.
\end{split}
\end{equation*}
\begin{proof}
See Appendix \ref{proof:tm3} for details. 
\end{proof}
\end{theorem}

\begin{corollary}
By setting $\omega_1 = ... = \omega_M = 1/M$, the following inequality holds under Theorem~\ref{tm:tm3}:
\begin{equation}
    \mathbb{E}[F(\tilde{\mathbf{x}}_T)] - F(\mathbf{x}_*) \leq \tilde{\mathcal{O}}\left(\frac{\sigma^2}{\mu M T \bar{K}}\right).
\end{equation}
\end{corollary}

\noindent\textbf{Remark}\quad Compared to FedNova \cite{wang2020tackling} that has convergence theory only for non-convex objectives, we have established the rigorous convergence theory for our method \texttt{FedaGrac} on strongly-convex objectives. Compared with Theorem \ref{tm:tm1}, \texttt{FedaGrac} not only converges to the optimal parameters, but also obtains a better convergence rate as $\tilde{\mathcal{O}}(1/\bar{K}) \leq \tilde{\mathcal{O}}(1/K_{\min})$. 

\section{Empirical Evaluation} \label{sec:experiment}

In this section, we conduct extensive experiments to evaluate the performance of \texttt{FedaGrac} in the real cases that are widely accepted by the existing studies. To further obtain an intuitive understanding of the numerical results, \texttt{FedaGrac} competes against other up-to-date benchmarks that are comparable under various settings. The code is implemented with PyTorch and available at \url{https://github.com/HarliWu/FedaGrac}. 

\subsection{Setup}

\begin{table}[!t]
    \centering
    \caption{Details for 2-layer CNN on Fashion-MNIST. Typically, Fashion-MNIST consists of grey-scale images possessing a single channel.}
    \resizebox{.47\textwidth}{!}{ 
    \begin{tabular}{ccccc}
    \hline
    Layer & Output Shape & Trainable Parameters & Activation & Hyperparameters \\ \hline
    
    Input & (1,28,28) & 0 & & \\ 
    Conv2d & (10, 24, 24) & 260 & ReLU & kernel size=5\\
    MaxPool2d & (10, 12, 12) & 0 & & kernel size=2\\
    Conv2d & (20, 8, 8) & 5020 & ReLU & kernel size=5\\
    Dropout2d & (20, 8, 8) & 0 & & p=0.5 \\
    MaxPool2d & (20, 4, 4) & 0 & & kernel size=2\\
    Flatten & 320 & 0 & & \\
    Dense & 50 & 16050 & ReLU & \\
    Dropout & 50 & 0 & & p=0.5 \\
    Dense & 10 & 510 & softmax & \\ \hline
    \end{tabular}
    }
    
    \label{tab:cnn}
\end{table}

\begin{figure*}[t]
    \centering
    \subfloat[AlexNet w/ step async] {\includegraphics[width=.24\textwidth]{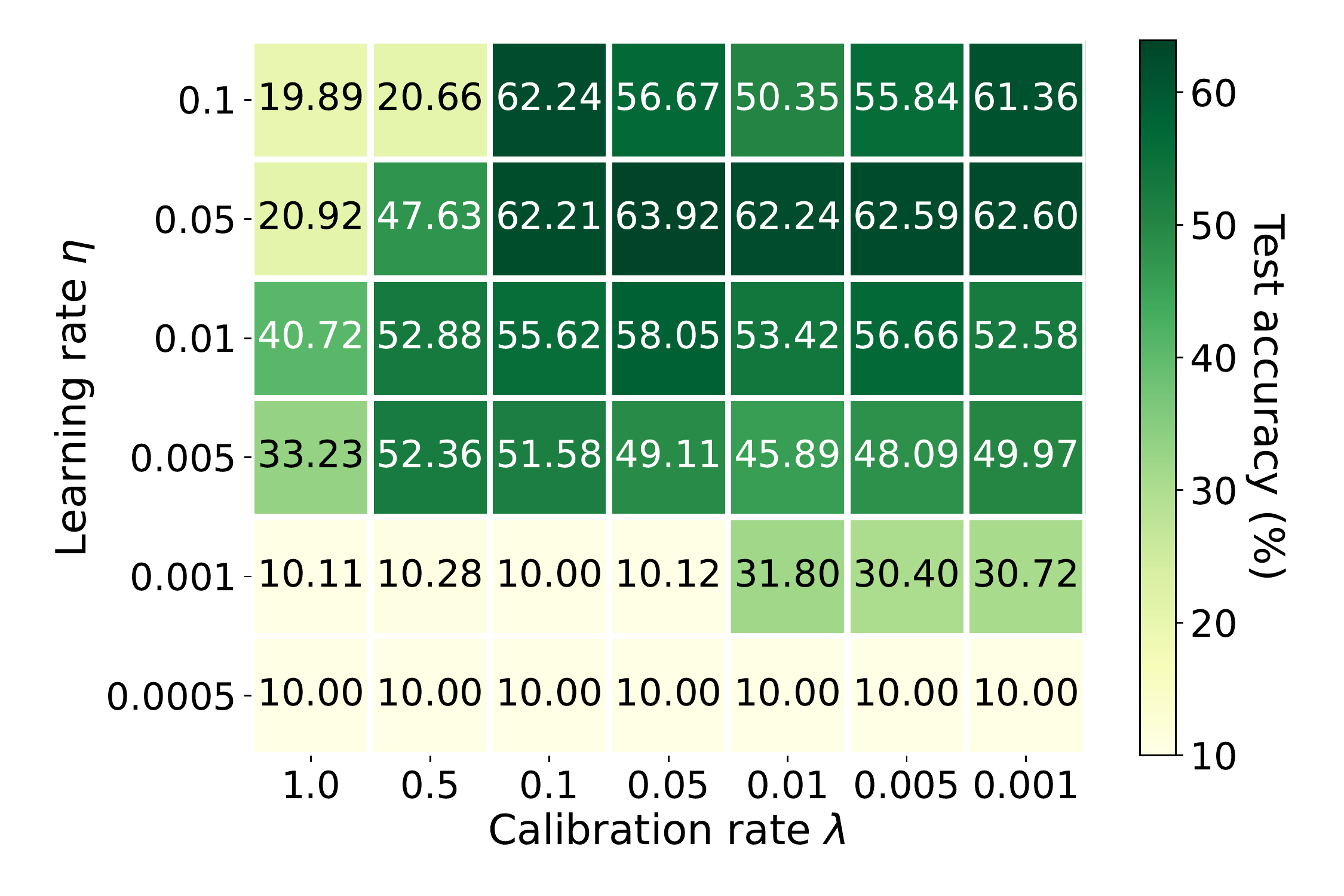} \label{fig:alexnet_step_acc_lam_lr}}
    \subfloat[LR w/ step async] {\includegraphics[width=.24\textwidth]{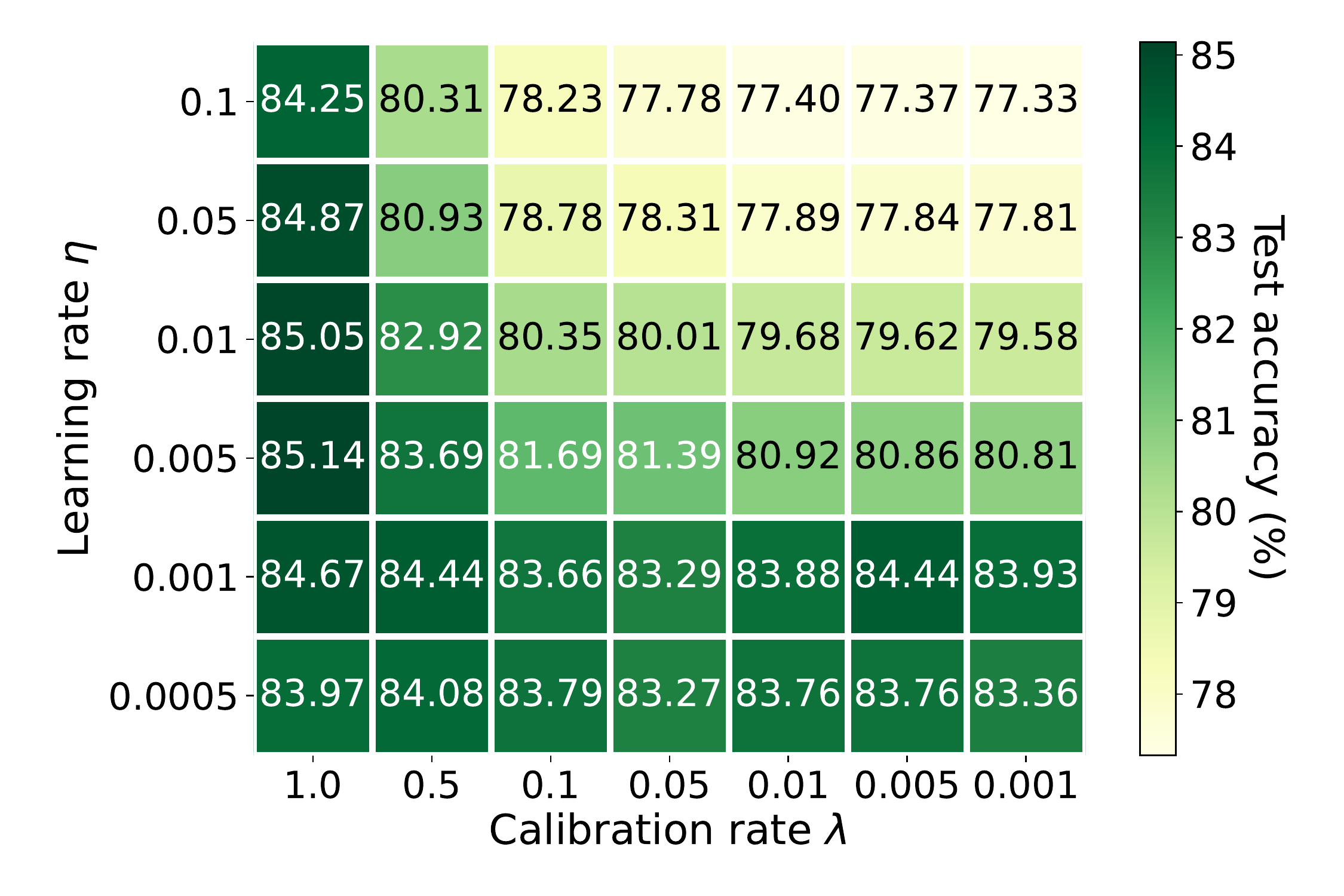} \label{fig:lr_step_acc_lam_lr}}
    \subfloat[AlexNet w/o step async] {\includegraphics[width=.24\textwidth]{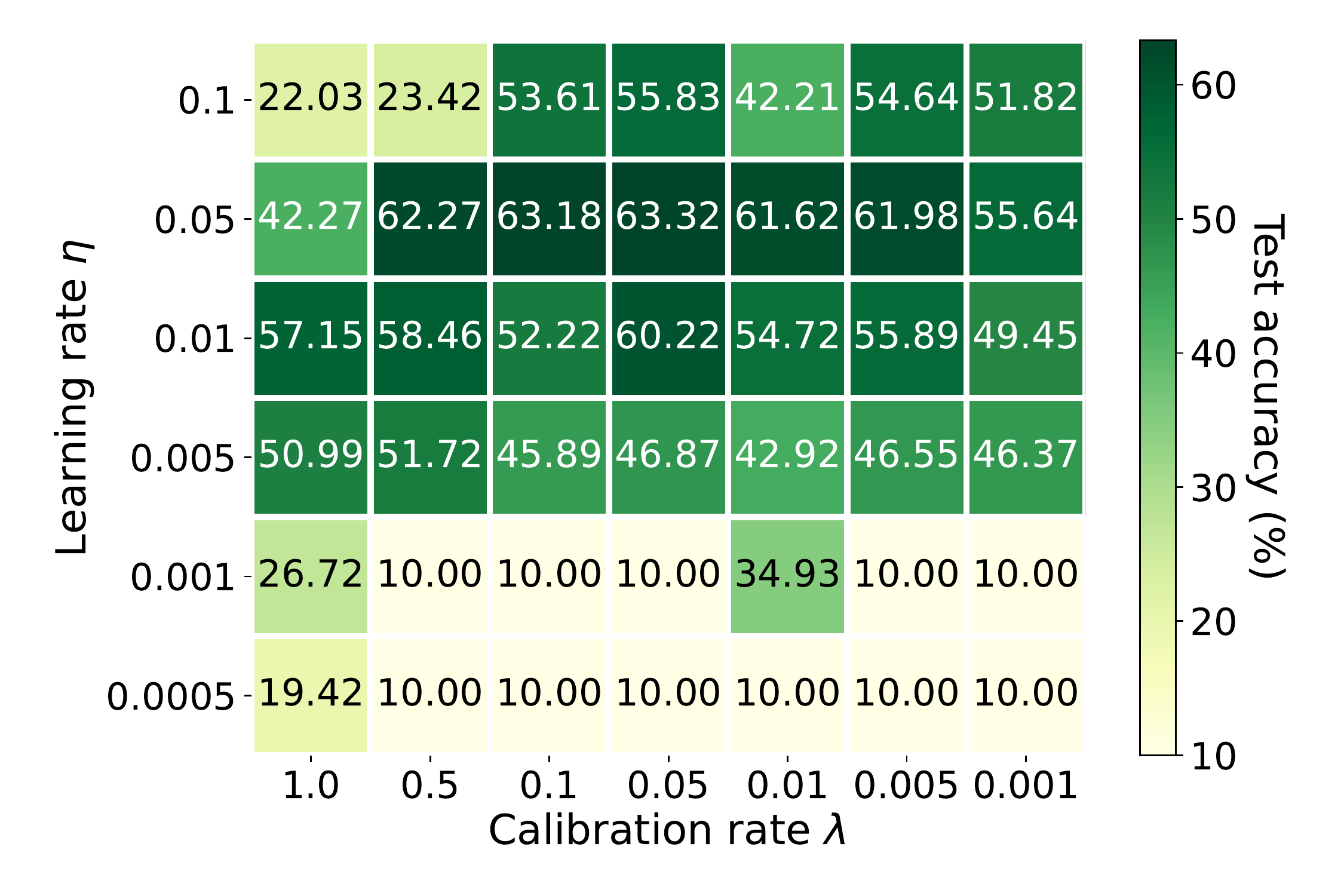} \label{fig:alexnet_nonstep_acc_lam_lr}}
    \subfloat[LR w/o step async] {\includegraphics[width=.24\textwidth]{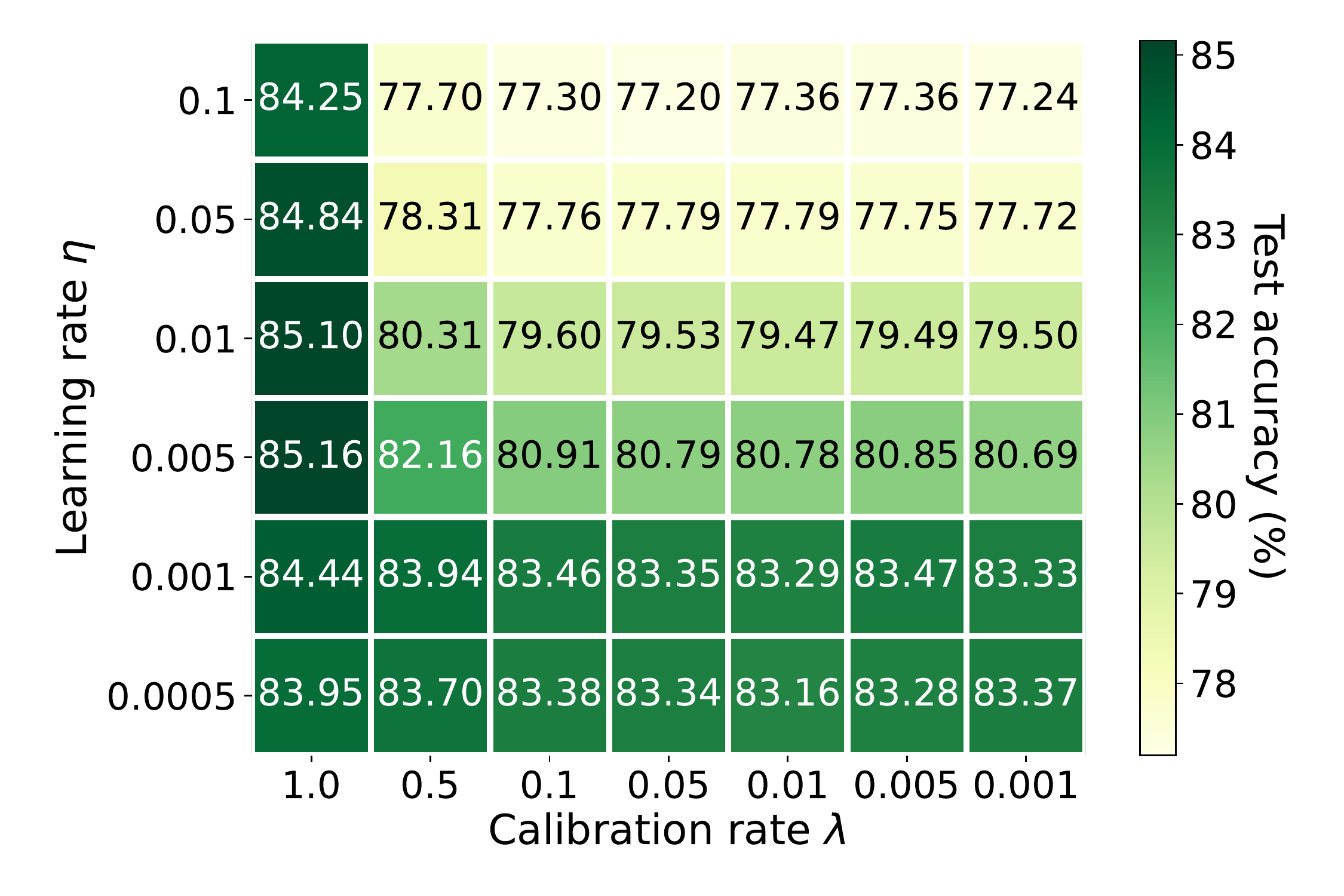} \label{fig:lr_nonstep_acc_lam_lr}}
    \caption{Comparison of various setting combinations for learning rate $\eta$ and calibration rate $\lambda$ using DP1 data distribution under AlexNet and LR after 100 communication rounds. The horizontal index indicates the value of $\lambda$ while the vertical index shows the value of $\eta$. The numeric in the box presents the averaged test accuracy of the last 10 rounds under the specific hyperparameter settings. The mean number of local updates is 500, and the variance with step asynchronism is 10000.}
    \label{fig:lr_calr_relationship}
\end{figure*}

\textbf{Datasets.} We leverage Fashion-MNIST \cite{xiao2017fashionmnist} to run the preliminary experiments in the previous sections. This dataset comprises 60000 28$\times$28 grey-scale training images and 10000 test images, which can be categorized into ten classes related to the clothes type. In this section, we utilize two more datasets: a9a\footnote{\url{https://www.csie.ntu.edu.tw/~cjlin/libsvmtools/datasets/}} and CIFAR-10 \cite{krizhevsky2009learning}. As a binary classification task, a9a consists of 32561 training samples and 16281 test samples, and each sample possesses 123 features. CIFAR-10 is a 10-category image classification task, constituting 60000 32$\times$32 color images divided into the training and test set with the size of 50000 and 10000, respectively. 

% For the assessment of convex objectives, we train a LR model with the handwritten digit database MNIST \cite{lecun1998gradient}, which possesses 60000 training examples and 10000 test examples with the size of $28 \times 28$. In addition, we investigate the performance under non-convex objectives through a image classification task CIFAR-10 \cite{krizhevsky2009learning} with VGG-19 \cite{vgg19}, a deep neural network with a total parameters of 20.03M \cite{zhu2021gradinit}. The detailed description is mentioned in Table \ref{tab:vgg}. 

\begin{table}[!t]
    \centering
    \caption{Details for AlexNet on CIFAR-10. Output shape follows the format of (channel, height, width). Generally, color images like CIFAR-10 dataset are with three channels.}
    \resizebox{.47\textwidth}{!}{ 
    \begin{tabular}{ccccc}
    \hline
    Layer & Output Shape & Trainable Parameters & Activation & Hyperparameters \\ \hline
    Input & (3,32,32) & 0 & & \\ 
    Conv2d & (64, 8, 8) & 23296 & ReLU & kernel size=11, stride=4, padding=5\\
    MaxPool2d & (64, 4, 4) & 0 & & kernel size=2, stride=2 \\
    Conv2d & (192, 4, 4) & 307392 & ReLU & kernel size=5, padding=2\\
    MaxPool2d & (192, 2, 2) & 0 & & kernel size=2, stride=2\\
    Conv2d & (384, 2, 2) & 663936 & ReLU & kernel size=3, padding=1\\
    Conv2d & (256, 2, 2) & 884992 & ReLU & kernel size=3, padding=1\\
    Conv2d & (256, 2, 2) & 590080 & ReLU & kernel size=3, padding=1\\
    MaxPool2d & (256, 1, 1) & 0 & & kernel size=2, stride=2\\
    Flatten & 256 & 0 & & \\
    Dropout & 256 & 0 & & p = 0.5 \\
    Dense & 2048 & 526336 & ReLU & \\
    Dropout & 2048 & 0 & & p = 0.5 \\
    Dense & 2048 & 4196352 & ReLU & \\ 
    Dense & 10 & 20490 & softmax & \\ \hline
    \end{tabular}
    }
    
    \label{tab:alexnet}
\end{table}

\begin{table}[!t]
    \centering
    \caption{Network architecture for VGG-19 on CIFAR-10.}
    \resizebox{.47\textwidth}{!}{ 
    \begin{tabular}{ccccc}
    \hline
    Layer & Output Shape & Trainable Parameters & Activation & Hyperparameters \\ \hline
    Input & (3,32,32) & 0 & & \\ 
    2 $\times$ Conv2d & (64, 32, 32) & 38720 & ReLU &  kernel size=3; padding=1 \\
    MaxPool2d & (64, 16, 16) & 0 & & kernel size=2, stride=2 \\
    2 $\times$ Conv2d & (128, 16, 16) & 221440 & ReLU &  kernel size=3; padding=1 \\
    MaxPool2d & (128, 8, 8) & 0 & & kernel size=2, stride=2 \\
    4 $\times$ Conv2d & (256, 8, 8) & 2065408 & ReLU &  kernel size=3; padding=1 \\
    MaxPool2d & (256, 4, 4) & 0 & & kernel size=2, stride=2 \\
    4 $\times$ Conv2d & (512, 4, 4) & 8259584 & ReLU &  kernel size=3; padding=1 \\
    MaxPool2d & (512, 2, 2) & 0 & & kernel size=2, stride=2 \\
    4 $\times$ Conv2d & (512, 2, 2) & 9439232 & ReLU &  kernel size=3; padding=1 \\
    MaxPool2d & (512, 1, 1) & 0 & & kernel size=2, stride=2 \\
    Flatten & 512 & 0 & & \\
    Dropout & 512 & 0 & & p = 0.5\\
    Dense & 512 & 262656 & ReLU & \\
    Dropout & 512 & 0 & & p = 0.5 \\
    Dense & 512 & 262656 & ReLU & \\
    Dense & 10 & 5130 & softmax & \\ \hline
    \end{tabular}
    }
    
    \label{tab:vgg}
\end{table}

\noindent \textbf{Models.} For the assessment of convex objectives, we train a logistic regression (LR) model using a9a. In addition, we investigate the performance under non-convex objectives through an image classification task CIFAR-10 \cite{krizhevsky2009learning} with AlexNet \cite{krizhevsky2012imagenet} and VGG-19 \cite{vgg19}, deep neural networks with total parameters of 7.21M and 20.55M, respectively. As for Fashion-MNIST, 2-layer CNN and LR are utilized to evaluate the performance under non-convex and convex objectives, respectively. Based on the dataset used, the details for 2-layer CNN, AlexNet and VGG-19 are separately described in Table \ref{tab:cnn}, Table \ref{tab:alexnet} and Table \ref{tab:vgg}. 

\noindent \textbf{Data Heterogeneity.} 
% We equally divide the whole dataset into $M$ subsets for each client in the i.i.d. settings. 
% As for the non-i.i.d. settings, we adopt two different partitioned ways. A traditional way is to disjoint the dataset via sharding, and thus each client holds at least $x$ classes, where $x$ is an integer and always smaller than the number of available categories. We let such a method be DP2-$x$ and ensure clients carry the same volume of data. However, the division approach is not applicable for a9a because \revise{it is a binary classification challenge}. Instead, we split the dataset across the clients following Dirichlet distribution with parameter 0.3, denoted as DP1. Such a partition is also available for CIFAR-10. 
As for the non-i.i.d. settings, we adopt two different partitioned ways. The first one that we split the dataset across the clients follows the Dirichlet distribution with parameter 0.3, denoted as DP1. This approach is suitable for both datasets. The other method disjoints the dataset via sharding, and thus each client holds 5 classes. We let such a method be DP2 and ensure clients carry the same volume of data. It is worth noting that this partition is only compatible with CIFAR-10 because a9a is a binary classification challenge. 
% A traditional way is to disjoint the dataset via sharding, and thus each client holds 5 classes, where $x$ is an integer and always smaller than the number of available categories. We let such a method be DP2-$x$ and ensure clients carry the same volume of data. However, the division approach is not applicable for a9a because \revise{it is a binary classification challenge}. Instead, we split the dataset across the clients following Dirichlet distribution with parameter 0.3, denoted as DP1. Such a partition is also available for CIFAR-10. 

\noindent \textbf{Computational Heterogeneity.} To simulate a heterogeneous computing environment, we suppose the computation differences among workers follow the Gaussian distribution. Then, the number of local updates varies among clients and follows the normal distribution with predefined mean and variance. And the number of local updates may change over time for each client. 
% and subject to the real-time computational capacity. 
% There are two different ways to ensure the number of local iterations, namely, \textit{linear} and \textit{Gaussian}. For Client $i \in \{1, ..., M\}$, the number of local iterations determined by linear follows a mathematical formula that is $(100+150 \cdot i)$. Without special annotation, the default setting for \textit{Gaussian} is that the numbers of local updates among clients follow Gaussian distribution with the mean of 500 and the standard deviation of 10000. 

\noindent \textbf{Implementation and Hyperparameter Settings.} 
The experiments are conducted with an MPI-supported cluster with the configurations of 100GB RAM, 25 CPU cores, and 1 Nvidia P100 GPU. Based on the resource, we utilize 20 cores to act as clients and a single core as the federated server. Besides, the batch sizes throughout our experiments are set as 25 and 20 for CIFAR-10 and a9a, respectively. \revise{We choose FedAvg \cite{mcmahan2017communication}, FedNova \cite{wang2020tackling}, SCAFFOLD \cite{karimireddy2020scaffold} and FedProx \cite{li2020federated} as benchmarks and present the effectiveness of our proposed approach \texttt{FedaGrac}. For a fair comparison, we compare these algorithms with the results when they achieve the best performance under the constant learning rates $\{0.01, 0.008, 0.005\}$ and $\{0.005, 0.001, 0.0005\}$ for AlexNet/VGG-19 and LR, respectively. And other required hyperparameters are also carefully picked from a set, such as the coefficient of the regularization term for FedProx in $\{1, 0.1, 0.01\}$. We specified other unmentioned but necessary settings in the captions of the figures and the tables.}
 % \shiqi{font too small}
% The experiments are conducted with an MPI-supported node with the configurations of 100GB RAM, 25 CPU cores, and 1 Nvidia P100 GPU. Based on the resource, we utilize 20 cores to act as clients and a single core as the federated server. Besides, the batch size throughout all of our experiments is set to be 25 and 20 for CIFAR-10 and a9a, respectively. To obtain the best the results, the algorithms (i.e., \texttt{FedaGrac}, FedAvg \cite{mcmahan2017communication}, FedNova \cite{wang2020tackling}, SCAFFOLD \cite{karimireddy2020scaffold} and FedProx \cite{li2020federated}) pick one of the constant learning rates from \{0.03, \revise{0.02}, 0.01, 0.008\} for non-convex objectives and \{0.008, 0.005, 0.001\} for convex cases. The calibration rate $\lambda$ for \texttt{FedaGrac} is set to be \revise{0.03} and 1.0 for non-convex and convex objectives, respectively. Other settings like the mean and the variance of the number of local updates will be specified in the captions of the figures and the tables. 

% Besides, to obtain the best results for each algorithm, we set the learning rate as 0.05 for SCAFFOLD \cite{karimireddy2020scaffold} and \texttt{FedaGrac} while FedNova \cite{wang2020tackling}, FedAvg \cite{mcmahan2017communication, li2019convergence} and FedProx \cite{li2020federated} to be 0.01 under both VGG-19 and LR. Without annotation, for \texttt{FedaGrac}, we set the value of calibration rate $\lambda$ to be 0.05 for VGG-19, and the coefficient of proximal term for FedProx to be 0.5. 

\subsection{Numerical Results} 

% \makecell{\progressbar[width=1.6cm, ticksheight=0, linecolor=black, filledcolor=black]{0.01}}
\begin{table*}[]
\centering
\caption{The number of communication rounds when \revise{first} achieving the target test accuracy
% with DP2-5 (or the test accuracy of 25\% with DP2-2) 
under AlexNet and VGG-19. The computational capabilities among workers follow the Gaussian distribution with a mean of 500 and different variances (i.e., V = 0, V = 100, and V = 10000) using two different data distributions (i.e., DP1 and DP2). Random mode indicates the number of local updates on a client varies among communication rounds, while fixed mode does not possess the feature. Each experiment runs for a maximum of 200 rounds. }
\label{table:variance_data}
\renewcommand{\arraystretch}{1.5}
% \vspace{10px}
\resizebox{\textwidth}{!}{%
\begin{tabular}{cccccccccc}
\hline
 \multirow{2}{*}{Model}&\multirow{2}{*}{\makecell{Data\\Distribution}} &\multirow{2}{*}{\makecell{Target\\Accuracy}} & \multirow{2}{*}{Variance} & \multirow{2}{*}{Mode} & \multicolumn{4}{c}{Number of communication rounds ($\downarrow$)} \\ \cline{6-10}
 & & & & & \texttt{FedaGrac} & FedAvg & FedNova & SCAFFOLD & FedProx \\ \cline{1-10}
%  \multirow{5}{*}{DP2-5} & V = 0 & - & 45.99\textcolor[RGB]{96,96,96}{$\pm${0.24}} & 50.82\textcolor[RGB]{96,96,96}{$\pm${0.27}} & 42.72\textcolor[RGB]{96,96,96}{$\pm${0.18}} &  50.01\textcolor[RGB]{96,96,96}{$\pm${0.15}} \\ \cline{2-7}

 \multirow{5}{*}{AlexNet}&\multirow{5}{*}{DP1} &\multirow{5}{*}{68\%} & V = 0 & - & \makecell[l]{\progressbar[width=1.7cm, ticksheight=0, linecolor=black, filledcolor=black]{0.56}} \quad \textbf{113} & \makecell[l]{\progressbar[width=1.7cm, ticksheight=0, linecolor=black, filledcolor=black]{0.61}} \quad 123 & \makecell[l]{\progressbar[width=1.7cm, ticksheight=0, linecolor=black, filledcolor=black]{0.64}} \quad 127 &  \makecell[l]{\progressbar[width=1.7cm, ticksheight=0, linecolor=black, filledcolor=black]{0.56}} \quad \textbf{113} &  \makecell[l]{\progressbar[width=1.7cm, ticksheight=0, linecolor=black, filledcolor=black]{0.72}} \quad 145 \\ \cline{4-10} 
 &&& \multirow{2}{*}{V = 100} & fixed & \makecell[l]{\progressbar[width=1.7cm, ticksheight=0, linecolor=black, filledcolor=black]{0.53}} \quad \textbf{106} & \makecell[l]{\progressbar[width=1.7cm, ticksheight=0, linecolor=black, filledcolor=black]{0.65}} \quad 130 & \makecell[l]{\progressbar[width=1.7cm, ticksheight=0, linecolor=black, filledcolor=black]{0.73}} \quad 147 &  \makecell[l]{\progressbar[width=1.7cm, ticksheight=0, linecolor=black, filledcolor=black]{0.57}} \quad 114 &  \makecell[l]{\progressbar[width=1.7cm, ticksheight=0, linecolor=black, filledcolor=black]{0.72}} \quad 144 \\ \cline{5-10} 
 && && random & \makecell[l]{\progressbar[width=1.7cm, ticksheight=0, linecolor=black, filledcolor=black]{0.58}} \quad \textbf{116} & \makecell[l]{\progressbar[width=1.7cm, ticksheight=0, linecolor=black, filledcolor=black]{0.7}} \quad 140 & \makecell[l]{\progressbar[width=1.7cm, ticksheight=0, linecolor=black, filledcolor=black]{0.77}} \quad 154 &  \makecell[l]{\progressbar[width=1.7cm, ticksheight=0, linecolor=black, filledcolor=black]{0.67}} \quad 133 &  \makecell[l]{\progressbar[width=1.7cm, ticksheight=0, linecolor=black, filledcolor=black]{0.70}} \quad 140\\ \cline{4-10} 
 &&& \multirow{2}{*}{V = 10000} & fixed & \makecell[l]{\progressbar[width=1.7cm, ticksheight=0, linecolor=black, filledcolor=black]{0.63}} \quad \textbf{126} & \makecell[l]{\progressbar[width=1.7cm, ticksheight=0, linecolor=black, filledcolor=black]{0.78}} \quad 156 & \makecell[l]{\progressbar[width=1.7cm, ticksheight=0, linecolor=black, filledcolor=black]{0.86}} \quad 172 &  \makecell[l]{\progressbar[width=1.7cm, ticksheight=0, linecolor=black, filledcolor=black]{0.70}} \quad 141 &  \makecell[l]{\progressbar[width=1.7cm, ticksheight=0, linecolor=black, filledcolor=black]{0.71}} \quad 142 \\ \cline{5-10}
 && && random & \makecell[l]{\progressbar[width=1.7cm, ticksheight=0, linecolor=black, filledcolor=black]{0.60}} \quad \textbf{121} & \makecell[l]{\progressbar[width=1.7cm, ticksheight=0, linecolor=black, filledcolor=black]{0.88}} \quad 177 & \makecell[l]{\progressbar[width=1.7cm, ticksheight=0, linecolor=black, filledcolor=black]{0.85}} \quad 170 & \makecell[l]{\progressbar[width=1.7cm, ticksheight=0, linecolor=black, filledcolor=black]{0.68}} \quad 136 & \makecell[l]{\progressbar[width=1.7cm, ticksheight=0, linecolor=black, filledcolor=black]{0.76}} \quad 152 \\ \hline
 
 \multirow{5}{*}{AlexNet} & \multirow{5}{*}{DP2} &\multirow{5}{*}{70\%} & V = 0 & - & \makecell[l]{\progressbar[width=1.7cm, ticksheight=0, linecolor=black, filledcolor=black]{0.76}} \quad 152 & \makecell[l]{\progressbar[width=1.7cm, ticksheight=0, linecolor=black, filledcolor=black]{0.92}} \quad 183 & \makecell[l]{\progressbar[width=1.7cm, ticksheight=0, linecolor=black, filledcolor=black]{0.93}} \quad 186 &  \makecell[l]{\progressbar[width=1.7cm, ticksheight=0, linecolor=black, filledcolor=black]{0.80}} \quad 160 & \makecell[l]{\progressbar[width=1.7cm, ticksheight=0, linecolor=black, filledcolor=black]{0.73}} \quad \textbf{147} \\ \cline{4-10} 
 &&& \multirow{2}{*}{V = 100} & fixed & \makecell[l]{\progressbar[width=1.7cm, ticksheight=0, linecolor=black, filledcolor=black]{0.56}} \quad \textbf{111} & \makecell[l]{\progressbar[width=1.7cm, ticksheight=0, linecolor=black, filledcolor=black]{0.90}} \quad 179 & \makecell[l]{\progressbar[width=1.7cm, ticksheight=0, linecolor=black, filledcolor=black]{0.59}} \quad 119 &  \makecell[l]{\progressbar[width=1.7cm, ticksheight=0, linecolor=black, filledcolor=black]{0.62}} \quad 124 &  \makecell[l]{\progressbar[width=1.7cm, ticksheight=0, linecolor=black, filledcolor=black]{0.76}} \quad 143 \\ \cline{5-10} 
 && && random & \makecell[l]{\progressbar[width=1.7cm, ticksheight=0, linecolor=black, filledcolor=black]{0.56}} \quad \textbf{112} & \makecell[l]{\progressbar[width=1.7cm, ticksheight=0, linecolor=black, filledcolor=black]{1.00}} \quad 200+ & \makecell[l]{\progressbar[width=1.7cm, ticksheight=0, linecolor=black, filledcolor=black]{0.97}} \quad 195 &  \makecell[l]{\progressbar[width=1.7cm, ticksheight=0, linecolor=black, filledcolor=black]{0.69}} \quad 137 & \makecell[l]{\progressbar[width=1.7cm, ticksheight=0, linecolor=black, filledcolor=black]{0.70}} \quad 141\\ \cline{4-10} 
 &&& \multirow{2}{*}{V = 10000} & fixed & \makecell[l]{\progressbar[width=1.7cm, ticksheight=0, linecolor=black, filledcolor=black]{0.56}} \quad \textbf{111} & \makecell[l]{\progressbar[width=1.7cm, ticksheight=0, linecolor=black, filledcolor=black]{1.00}} \quad 200+ & \makecell[l]{\progressbar[width=1.7cm, ticksheight=0, linecolor=black, filledcolor=black]{0.56}} \quad 113 &  \makecell[l]{\progressbar[width=1.7cm, ticksheight=0, linecolor=black, filledcolor=black]{0.66}} \quad 131 & \makecell[l]{\progressbar[width=1.7cm, ticksheight=0, linecolor=black, filledcolor=black]{0.72}} \quad 145 \\ \cline{5-10}
 && && random & \makecell[l]{\progressbar[width=1.7cm, ticksheight=0, linecolor=black, filledcolor=black]{0.59}} \quad \textbf{118} & \makecell[l]{\progressbar[width=1.7cm, ticksheight=0, linecolor=black, filledcolor=black]{1.0}} \quad 200+ &  \makecell[l]{\progressbar[width=1.7cm, ticksheight=0, linecolor=black, filledcolor=black]{1.0}} \quad 200+ &  \makecell[l]{\progressbar[width=1.7cm, ticksheight=0, linecolor=black, filledcolor=black]{0.61}} \quad 123 &  \makecell[l]{\progressbar[width=1.7cm, ticksheight=0, linecolor=black, filledcolor=black]{0.76}} \quad 152 \\ \hline
 
 \multirow{5}{*}{VGG-19} & \multirow{5}{*}{DP2}&\multirow{5}{*}{80\%} & V = 0 & - & \makecell[l]{\progressbar[width=1.7cm, ticksheight=0, linecolor=black, filledcolor=black]{0.36}} \quad 73 & \makecell[l]{\progressbar[width=1.7cm, ticksheight=0, linecolor=black, filledcolor=black]{0.41}} \quad 83 & \makecell[l]{\progressbar[width=1.7cm, ticksheight=0, linecolor=black, filledcolor=black]{0.39}} \quad 79 &  \makecell[l]{\progressbar[width=1.7cm, ticksheight=0, linecolor=black, filledcolor=black]{0.36}} \quad \textbf{72} &  \makecell[l]{\progressbar[width=1.7cm, ticksheight=0, linecolor=black, filledcolor=black]{0.45
 }} \quad 90 \\ \cline{4-10} 
 &&& \multirow{2}{*}{V = 100} & fixed & \makecell[l]{\progressbar[width=1.7cm, ticksheight=0, linecolor=black, filledcolor=black]{0.36}} \quad 73 & \makecell[l]{\progressbar[width=1.7cm, ticksheight=0, linecolor=black, filledcolor=black]{0.37}} \quad 75 & \makecell[l]{\progressbar[width=1.7cm, ticksheight=0, linecolor=black, filledcolor=black]{0.36}} \quad 72 &  \makecell[l]{\progressbar[width=1.7cm, ticksheight=0, linecolor=black, filledcolor=black]{0.33}} \quad \textbf{66} &  \makecell[l]{\progressbar[width=1.7cm, ticksheight=0, linecolor=black, filledcolor=black]{0.36}} \quad 72 \\ \cline{5-10} 
 && && random & \makecell[l]{\progressbar[width=1.7cm, ticksheight=0, linecolor=black, filledcolor=black]{0.36}} \quad \textbf{73} & \makecell[l]{\progressbar[width=1.7cm, ticksheight=0, linecolor=black, filledcolor=black]{0.42}} \quad 85 & \makecell[l]{\progressbar[width=1.7cm, ticksheight=0, linecolor=black, filledcolor=black]{0.37}} \quad 74 &  \makecell[l]{\progressbar[width=1.7cm, ticksheight=0, linecolor=black, filledcolor=black]{0.39}} \quad 78 &  \makecell[l]{\progressbar[width=1.7cm, ticksheight=0, linecolor=black, filledcolor=black]{0.51}} \quad 102\\ \cline{4-10} 
 &&& \multirow{2}{*}{V = 10000} & fixed & \makecell[l]{\progressbar[width=1.7cm, ticksheight=0, linecolor=black, filledcolor=black]{0.38}} \quad 77 & \makecell[l]{\progressbar[width=1.7cm, ticksheight=0, linecolor=black, filledcolor=black]{0.36}} \quad 73 & \makecell[l]{\progressbar[width=1.7cm, ticksheight=0, linecolor=black, filledcolor=black]{0.35}} \quad \textbf{70} &  \makecell[l]{\progressbar[width=1.7cm, ticksheight=0, linecolor=black, filledcolor=black]{0.36}} \quad 72 &  \makecell[l]{\progressbar[width=1.7cm, ticksheight=0, linecolor=black, filledcolor=black]{0.39}} \quad 77 \\ \cline{5-10}
 && && random & \makecell[l]{\progressbar[width=1.7cm, ticksheight=0, linecolor=black, filledcolor=black]{0.35}} \quad \textbf{71} & \makecell[l]{\progressbar[width=1.7cm, ticksheight=0, linecolor=black, filledcolor=black]{0.42}} \quad 85 &  \makecell[l]{\progressbar[width=1.7cm, ticksheight=0, linecolor=black, filledcolor=black]{0.38}} \quad 76 &  \makecell[l]{\progressbar[width=1.7cm, ticksheight=0, linecolor=black, filledcolor=black]{0.36}} \quad 72 &  \makecell[l]{\progressbar[width=1.7cm, ticksheight=0, linecolor=black, filledcolor=black]{0.49}} \quad 99 \\ \hline

\end{tabular}%
}
\vspace{-10px}
\label{tab1}
\end{table*}

\noindent \textbf{Performance under Various combinations for learning rate and calibration rate.} \revise{As learning rate $\eta$ and calibration rate $\lambda$ need tuning in \texttt{FedaGrac}, we first explore how to set both hyperparameters scientifically.} Figure \ref{fig:lr_calr_relationship} depicts the test accuracy under various relations between $\eta$ and $\lambda$. As we observe, the differences regarding the convexity are quite significant, e.g., AlexNet in Figure \ref{fig:alexnet_step_acc_lam_lr} and LR in Figure \ref{fig:lr_step_acc_lam_lr}, while the computation heterogeneity has minor influence on the selection of hyperparameters under the same model, e.g., AlexNet in Figure \ref{fig:alexnet_step_acc_lam_lr} and Figure \ref{fig:alexnet_nonstep_acc_lam_lr}. Based on the acquired results, we discuss how to set the hyperparameters for \texttt{FedaGrac} under convex or non-convex objectives. 

Both Figure \ref{fig:alexnet_step_acc_lam_lr} and Figure \ref{fig:alexnet_nonstep_acc_lam_lr} illustrate the performance under AlexNet with and without computational heterogeneity. In both cases, most $\lambda$s achieve the highest accuracy at $\eta=0.05$, while some have the best performance at $\eta=0.01$. When the learning rate initializes with a value smaller or equal to 0.001, most AlexNets seem untrained after 100 rounds because they are less likely to escape a saddle point. Although some portfolios successfully get out of the minima, they still cannot outperform the aforementioned settings because they may (i) trap into a non-optimal stable point or (ii) need a longer period to reach the optimal solution. A constant $\lambda$ that performs well in all learning rates does not exist. However, when we shrink the choice of learning rate between 0.01 and 0.05, $\lambda=0.05$ has a remarkable performance. \revise{In our experiments, the calibration rate is chosen from $\{0.01, ..., 0.05\}$ depending on the algorithm's performance.}

Figure \ref{fig:lr_step_acc_lam_lr} and Figure \ref{fig:lr_nonstep_acc_lam_lr} present the results under the convex objectives. Regardless of the step asynchronism, $\lambda = 1$ always has remarkable performance for any learning rate. And it is noticeable that \texttt{FedaGrac} can obtain the best performance when $\lambda = 1$ and $\eta = 0.005$. As for a $\lambda \neq 1$, \texttt{FedaGrac} can achieve better performance as the learning rate becomes smaller. \revise{With such a phenomenon, we hypothesize that \texttt{FedaGrac} cannot exactly reach the identical minimizer when $\lambda \neq 1$ and approaches the expected point as the learning rate reduces. }
% Compared to non-convex objectives, the reason is possible that the computed gradients are less noisy such that most dimensions of corrected gradients are consistent with the orientation towards the optimal solutions. As the learning rate becomes smaller, every calibration rate can achieve improved performance except $\lambda = 1$. However, these results do not exceed the accuracy under the setting of $\lambda = 1$ and $\eta = 0.005$. 
% global and local update orientations always point to the only optimal solutions with respect to the global and the local objectives, respectively
% describes how many communication rounds a model achieves the test accuracy of 50\% (or 25\% for DP2-2).

\noindent \textbf{Performance under various data distributions.} \revise{Table \ref{table:variance_data} validates our algorithm under different data heterogeneities, i.e., DP1 and DP2 under AlexNet.} The target accuracy is determined by the best performance that these five algorithms can achieve when they run a constant number of updates. By comparing each algorithm under these two data distributions, DP2 is more challenging for FedAvg and FedNova because the algorithms generally require more communication rounds to achieve the target. Even worse, these two algorithms cannot achieve the goal within 200 rounds in some DP2 settings. As for the regularization-based approach (i.e., FedProx) and the variance reduction approaches (i.e., \texttt{FedaGrac} and SCAFFOLD), the task shifting does not cause a distinct influence\footnote{The difference between the numbers of the communication rounds is less than 15\%.} in terms of the required communication rounds. As we can see in both cases with computational differences, \texttt{FedaGrac} demonstrates the superiority over other benchmarks. 
% Obviously, the number of communication rounds using DP2-5 is less than the one using DP2-2 for all baselines with the same computational capacities since DP2-2 is more challenging to train a generalization model.  \revise{Under a significant data heterogeneity (i.e., DP2-2), \texttt{FedaGrac} demonstrates indefeasible superiority over other benchmarks.} 
% Under the data setting of DP2-2, \texttt{FedaGrac} considerably outperforms all existing baselines. 
% As an approach targeting to solve computational heterogeneity, FedNova requires up to 2$\times$ more communication rounds than our proposed approach. \revise{When compared to the traditional algorithms (i.e., FedAvg, SCAFFOLD, and FedProx), \texttt{FedaGrac} costs at least 50\% less rounds to achieve 25\% test accuracy. Likewise, our approach is still able to obtain a well-performed AlexNet after fewer-round training under DP2-5 data distribution.}
% Also, in some cases, FedNova has degraded performance when compared to the traditional federated learning approaches, i.e., FedAvg and SCAFFOLD. 
% As for the comparison between two classical algorithms, the variance-reduced approach (i.e., SCAFFOLD) better fits the scenario. However, with the existence of step asynchronism, SCAFFOLD needs 2$\times$ more global synchronizations than \texttt{FedaGrac}. Similar conclusions can be drawn from DP2-5 as well, but the dominance of \texttt{FedaGrac} is not that distinct. 

\begin{figure*}[t]
\flushleft
% \subfloat[AlexNet with 500] {\includegraphics[width=.24\textwidth]{Figures/alexnet_loss_500.pdf} \label{fig:alexnet_loss_500}}
\subfloat[AlexNet with 500] {\includegraphics[width=.24\textwidth]{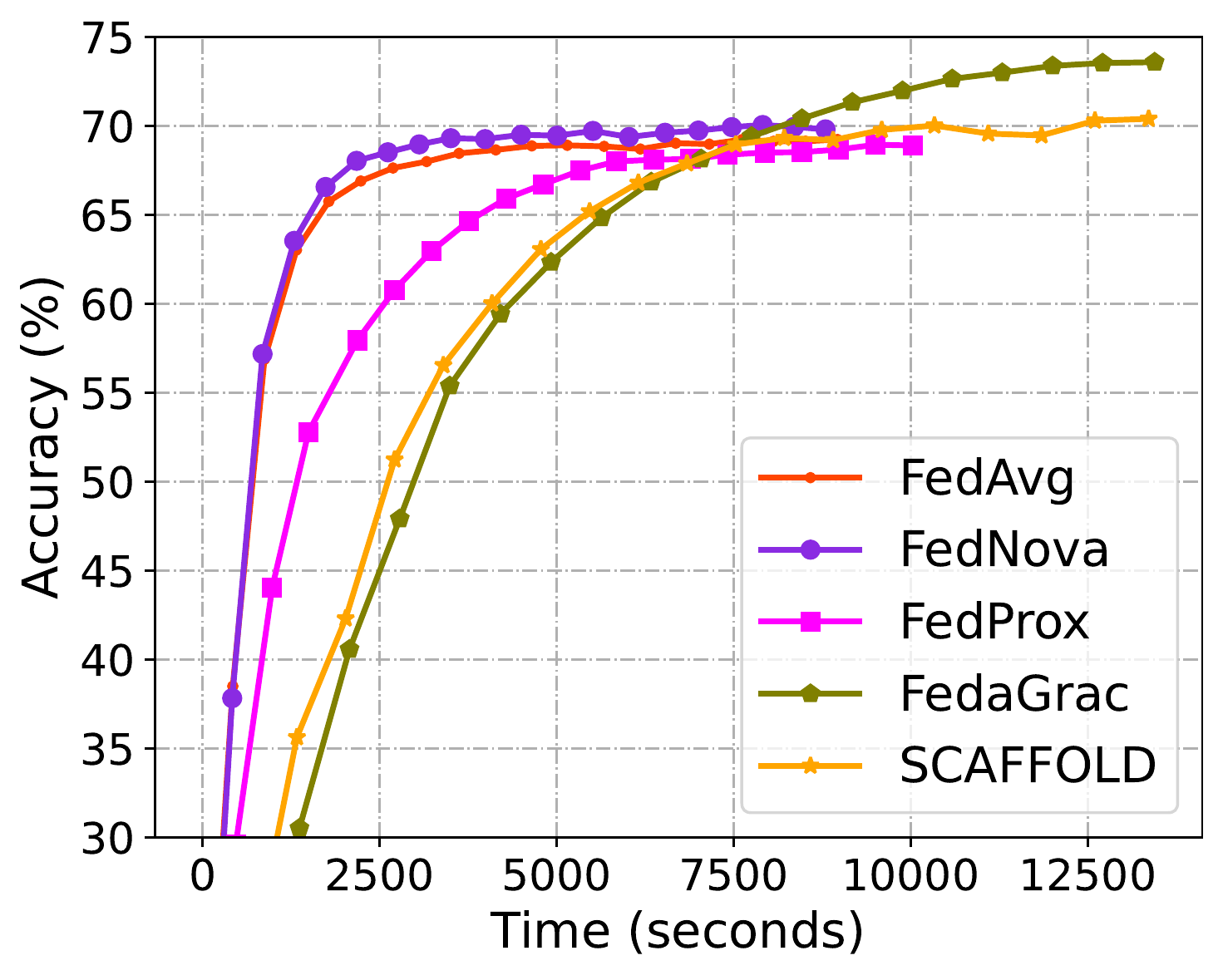} \label{fig:alexnet_acc_500}}
% \subfloat[AlexNet with 1000] {\includegraphics[width=.24\textwidth]{Figures/alexnet_loss_1000.pdf}\label{fig:alexnet_loss_1000}}
\subfloat[AlexNet with 1000] {\includegraphics[width=.24\textwidth]{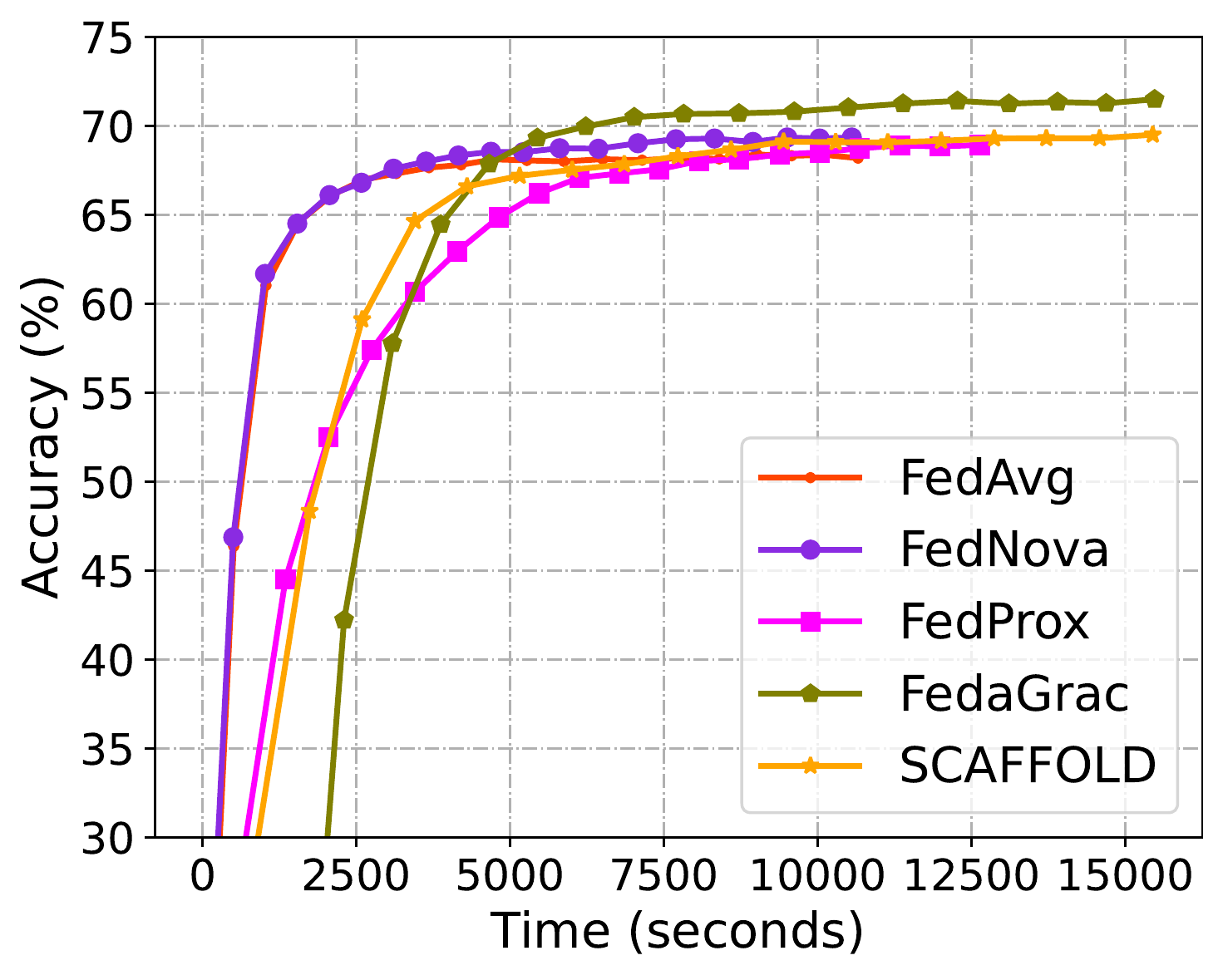}\label{fig:alexnet_acc_1000}}
% \\
% \subfloat[LR with 500] {\includegraphics[width=.24\textwidth]{Figures/lr_loss_500.pdf} \label{fig:lr_loss_500}}
\subfloat[LR with 500] {\includegraphics[width=.24\textwidth]{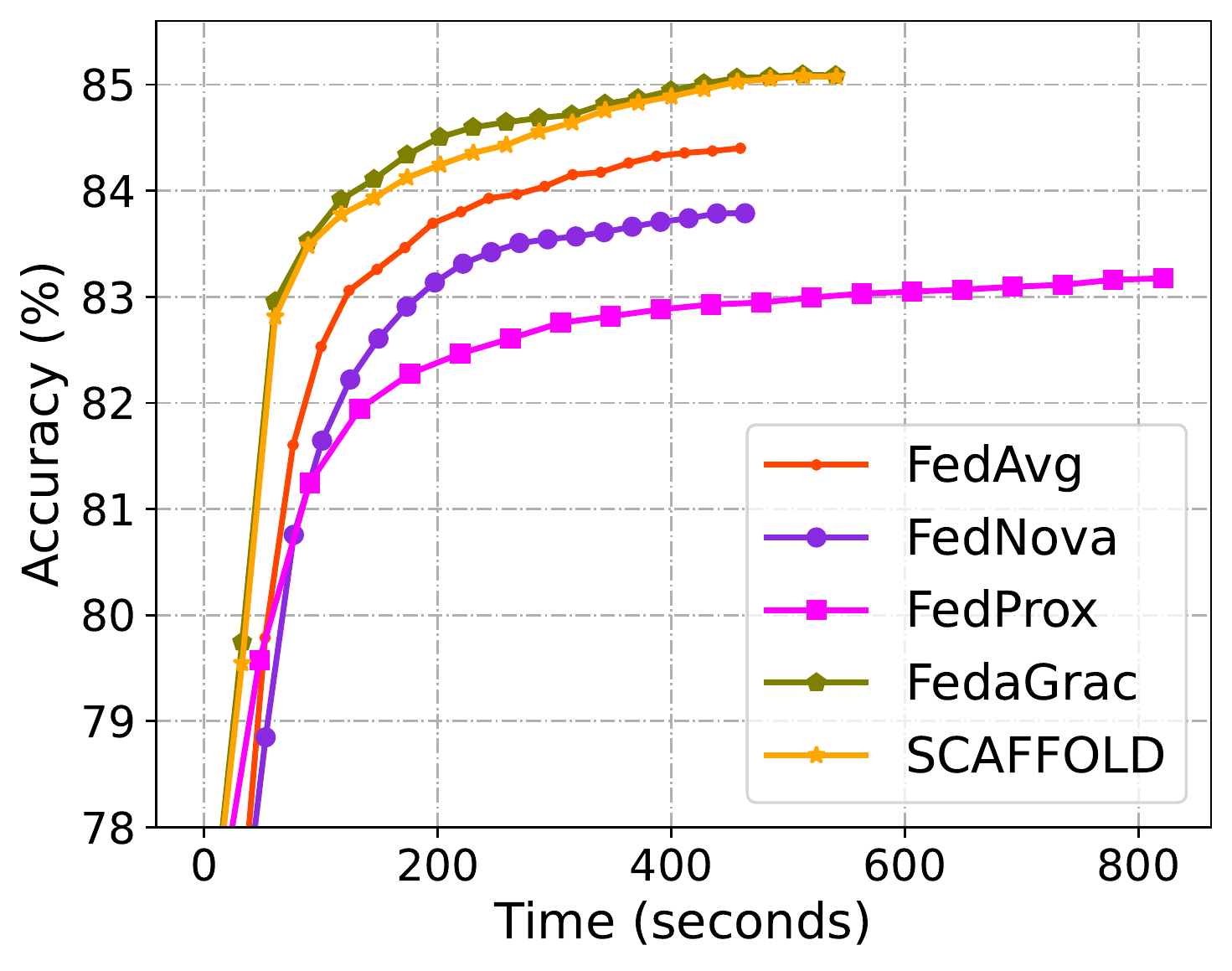} \label{fig:lr_acc_500}}
% \subfloat[LR with 1000] {\includegraphics[width=.24\textwidth]{Figures/lr_loss_1000.pdf}\label{fig:lr_loss_1000}}
\subfloat[LR with 1000] {\includegraphics[width=.24\textwidth]{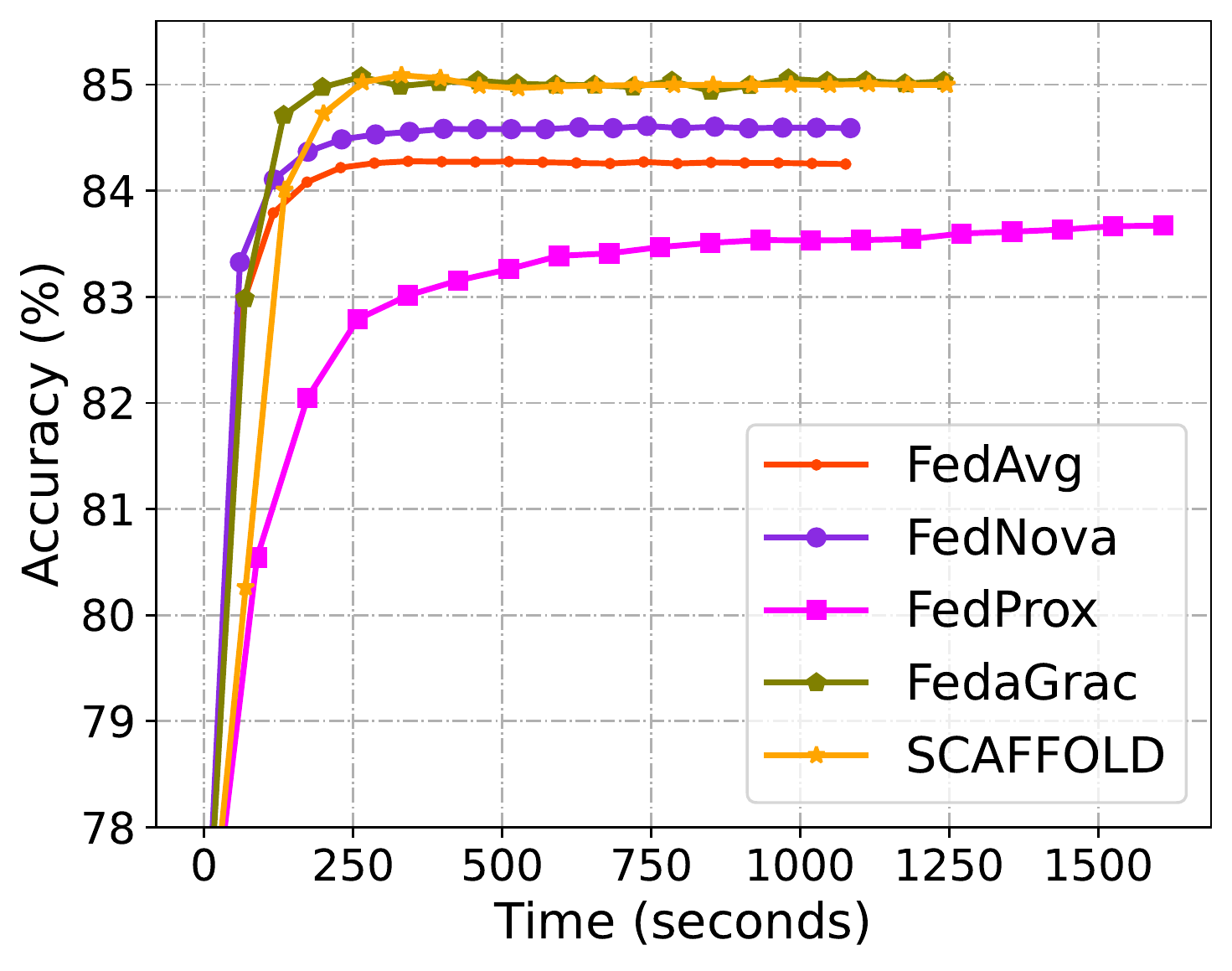}\label{fig:lr_acc_1000}}
\vspace{-5px}
\caption[t]{Test accuracy v.s. Training time under different gaussian mean w.r.t. the fixed variance of 10000. Leftmost two figures: AlexNet using DP2; Rightmost two figures: LR using DP1. The number of local updates on a client is fixed and initialized at the beginning of the model training. Each algorithm runs for a total of 200 rounds, and the gap between two markers represents an interval of 10 communication rounds. }
\vspace{-10px}
\label{fig:different_mean_alexnet_lr}
\end{figure*}

\noindent \revise{\textbf{Performance under various neural networks.} In addition to exploring various data distributions based on Table \ref{table:variance_data}, we investigate the performance of \texttt{FedaGrac} under different neural networks. As we notice, the approach in VGG-19 does not outperform all benchmarks in some computation heterogeneity cases. Specifically, it requires several more rounds than the best algorithm. } An explanation for this phenomenon is that obtaining an 80\%-accuracy VGG-19 on CIFAR-10 is not a difficult task. In contrast to getting an AlexNet with a test accuracy of 70\%, the algorithms can adopt a greater learning rate to improve training efficiency. Since there are some restricted terms in \texttt{FedaGrac}, it is reasonable that our proposed algorithm cannot outperform the benchmarks. Meanwhile, it is common that some benchmarks cannot outperform FedAvg \cite{li2021federated}. However, it is worth noting that, as presented in Figure \ref{fig:different_mean_alexnet_lr}, the faster algorithm may not surpass the slower ones in terms of the final test accuracy. 
% (As we know, FedNova is supposed to be equivalent to FedAvg when V=0. However, due to the normalize process, the accuracy is possibly lost, such that it is possible that they don't have equivalent results. )
% However, such differences are minor because the worst cases incur much greater gaps, e.g., FedProx. 

\noindent \textbf{Performance under various computational capabilities.} \revise{While adopting Gaussian distribution to tune the computation heterogeneity, we should manually set both mean and variance. To explore whether these two hyperparameters influence the algorithms' performances, we conduct extensive experiments, and the relevant results are presented in Table \ref{table:variance_data} and Figure \ref{fig:different_mean_alexnet_lr}.} Table \ref{table:variance_data} evaluates the performance under the computational capabilities with a constant mean of 500 and different variances, while Figure \ref{fig:different_mean_alexnet_lr} assesses the convergence tendency under a fixed variance of 10000 and diverse means. 

\revise{Table \ref{table:variance_data} presents the results given different variances with/without time-varying local updates. Our analysis is mainly based on AlexNet because it gives noticeable differences when the variances or the modes switch. } Admittedly, these algorithms are not sensitive to whether the number of local updates is time-varying. However, it is worth noting that FedNova is vulnerable to time-varying settings in DP2. This is because a large learning rate may lead FedNova to a surrogate solution \cite{mitra2021linear} such that FedNova has to adopt a smaller learning rate to achieve the target accuracy. In contrast to time-varying local updates, the variance plays an important role in training efficiency. When the variance becomes larger, it is likely that the algorithms require more communication rounds. Nevertheless, a greater variance sometimes improves the training efficiency of those algorithms which mitigate the client-drift effect, i.e., \texttt{FedaGrac}, SCAFFOLD, and FedProx. 

% can act as an accelerator to speed up the training performance, especially for variance-reduction approaches under a difficult task because they mitigate the client-drift effect. (avoiding client-drift deterioration)

% Our proposed algorithm is not sensitive to computational heterogeneity since the number of communication rounds does not suffer as dramatic changes as other benchmarks do. Besides, although a greater variance and time-varying local update can increase the training difficulty, the results cannot mirror the hardship. With the phenomenon, we hypothesize that all clients have locally reached a stable point before the model aggregation.

% all benchmarks under deep neural networks achieving an acceptable performance depend on how the averaged number of local updates fluctuates over time.

% We first discuss the results shown in Table \ref{table:variance_data}. Our proposed algorithm is not sensitive to computational heterogeneity because there are no distinct numeric differences in terms of global synchronizations under different variances and modes. As for FedNova, its performance presumably gets degraded as the variance increases and as the mode changes from fixed to random. Although FedAvg and SCAFFOLD are also sensitive to step asynchronism, their performance can either improve or deteriorate as the variance and/or the mode change. 

% \shiqi{what is "to implement 200 rounds?"}
% \vspace{-8px}
Figure \ref{fig:different_mean_alexnet_lr} illustrates the entire training progress, i.e., the test accuracy with respect to the training time and the communication rounds under both convex and non-convex objectives. 
Our proposed algorithm achieves competitive accuracy compared to other baselines, despite a slow start likely taking place because the calibration is yet to settle the client-drift effects properly in the beginning. Although FedAvg and FedNova require half communication overhead as our proposed algorithm does, they cannot keep dominant alongside the training. Use AlexNet as an example (Figure \ref{fig:alexnet_acc_500} and \ref{fig:alexnet_acc_1000}), and \texttt{FedaGrac} is capable of achieving the same performance with fewer rounds. In addition, it is interesting to see FedProx consuming more time to implement 200 rounds than FedAvg. A reasonable explanation for this phenomenon is that extra computation is required by the regularization terms. As the model gets larger, this effect becomes minor since the communication consumption asymptotically occupies most training time (compare between LR (Figure \ref{fig:lr_acc_500}) and AlexNet (Figure \ref{fig:alexnet_acc_500}) for this heuristic conclusion). 
As a convex objective, LR depicts the issue of objective inconsistency (the latter two plots in Figure \ref{fig:different_mean_alexnet_lr}). The performances of FedAvg, FedNova, and FedProx are much worse than \texttt{FedaGrac} and SCAFFOLD. With the increasing mean and the unchanged variance, the deterioration gets mitigation but cannot eliminate. As for the comparison between SCAFFOLD and our proposed method, the latter possesses dominance nearly all the time. 

\section{Conclusion} \label{sec:conclusion}
This paper introduces a new algorithm named \texttt{FedaGrac} to tackle the challenges of both statistical heterogeneity and computation heterogeneity in FL. By calibrating the local client deviations according to an estimated global orientation in each communication round, the negative effect of step asynchronism on model accuracy can be greatly mitigated, and the training process is remarkably accelerated. We establish the theoretical convergence rate of \texttt{FedaGrac}. The results imply that \texttt{FedaGrac} admits a faster convergence rate and has a better tolerance to computation heterogeneity than the state-of-the-art approachs. Extensive experiments are also conducted to validate the advantages of \texttt{FedaGrac}.

\section*{Acknowledgments}

The authors would like to thank Shiqi He for the useful discussion and the anonymous reviewers for their constructive comments. This research was supported by fundings from the Key-Area Research and Development Program of Guangdong Province (No. 2021B0101400003), Hong Kong RGC Research Impact Fund (No. R5060-19), Areas of Excellence Scheme (AoE/E-601/22-R), General Research Fund (No. 152203/20E, 152244/21E, 152169/22E), Shenzhen Science and Technology Innovation Commission (JCYJ20200109142008673), the National Natural Science Foundation of China (Grant 62102131), and Natural Science Foundation of Jiangsu Province (Grant BK20210361).

\ifCLASSOPTIONcaptionsoff
  \newpage
\fi

\bibliographystyle{IEEEtran}
\bibliography{manuscript}

\allowdisplaybreaks
\appendices

\section{Proof of Theorem \ref{tm:tm1}} \label{proof:tm1}

The following lemma describes the relationship among three different parameters under strongly-convex function: 
\begin{lemma} \label{lemma:1}
Under Assumption \ref{ass:1} and Assumption \ref{ass:2}, given $\mathbf{a}, \mathbf{b}, \mathbf{c} \in \mathbb{R}^d$, the following formula holds under the strongly-convex objectives $F$: 
\begin{equation}
    \langle\nabla F(\mathbf{a}), \mathbf{b} - \mathbf{c}\rangle \leq F(\mathbf{b}) - F(\mathbf{c}) - \frac{\mu}{4} \|\mathbf{b} - \mathbf{c}\|_2^2 + L\|\mathbf{a} - \mathbf{c}\|_2^2
\end{equation}
\begin{proof}
With $\mathbf{a}, \mathbf{b} \text{ and } \mathbf{c}$ that are within the domain of $F$, we can get the following inequalities that come from Assumption \ref{ass:1} and Assumption \ref{ass:2}, respectively: 

\begin{equation*}
    \innerproduct{\nabla F(\mathbf{a})}{\mathbf{b} - \mathbf{a}} \leq F(\mathbf{b}) - F(\mathbf{a}) + \frac{L}{2} \|\mathbf{a} - \mathbf{b}\|_2^2
\end{equation*}
\begin{equation*}
    \innerproduct{\nabla F(\mathbf{a})}{\mathbf{a} - \mathbf{c}} \leq F(\mathbf{a}) - F(\mathbf{c}) - \frac{\mu}{2} \|\mathbf{a} - \mathbf{c}\|_2^2
\end{equation*}
By a formula that 
\begin{equation*}
    \|\mathbf{a} - \mathbf{b}\|_2^2 \leq 2\|\mathbf{a} - \mathbf{c}\|_2^2 + 2\|\mathbf{b} - \mathbf{c}\|_2^2
\end{equation*}
we have: 
\begin{equation*}
    \innerproduct{\nabla F(\mathbf{a})}{\mathbf{b} - \mathbf{c}} \leq F(\mathbf{b}) - F(\mathbf{c}) - \frac{\mu}{4} \|\mathbf{b} - \mathbf{c}\|_2^2 + \frac{L+\mu}{2}\|\mathbf{a} - \mathbf{c}\|_2^2
\end{equation*}
The inequality holds when $L \geq \mu$. 
\end{proof}
\end{lemma}

The update rule for FedAvg under heterogeneous steps:
\begin{equation}
    \mathbf{x}_{t+1} = \mathbf{x}_{t} - \eta\sum_{i=1}^M \w_i\sum_{k=0}^{K_i-1}\nabla f_i\left(\lx{i}{t}{k},\varepsilon_k^{(i)}\right)
\end{equation}
Therefore, the bound established for $\E\norm{\mathbf{x}_{t+1}-\optx}$ should be:
\begin{align}
    &\qquad\E\norm{\mathbf{x}_{t+1}-\optxs}\nonumber\\
    &= \E\norm{\mathbf{x}_{t} - \eta\sum_{i=1}^M \w_i\sum_{k=0}^{K_i-1}\nabla f_i\left(\lx{i}{t}{k},\varepsilon_k^{(i)}\right) - \optx} \nonumber \\
    &= \E\norm{\mathbf{x}_{t}-\optxs} + \E\norm{\eta\sum_{i=1}^M \w_i\sum_{k=0}^{K_i-1}\nabla f_i\left(\lx{i}{t}{k},\varepsilon_k^{(i)}\right)} \nonumber \\
    &\quad - 2\E\innerproduct{\mathbf{x}_{t}-\optx}{\eta\sum_{i=1}^M \w_i\sum_{k=0}^{K_i-1}\nabla f_i\left(\lx{i}{t}{k},\varepsilon_k^{(i)}\right)} \nonumber\\
    &=\E\norm{\mathbf{x}_{t}-\optxs} + \eta^2\sum_{i=1}^M\w_i^2K_i\sigma^2\\
    &\quad + \underbrace{\eta^2\E\norm{\sum_{i=1}^M \sum_{k=0}^{K_i-1}\w_i\gradthree{i}{t}{k}}}_{\mathcal{A}_2}\\ 
    &\quad \underbrace{- 2\E\innerproduct{\mathbf{x}_{t}-\optx}{\eta\sum_{i=1}^M \w_i\sum_{k=0}^{K_i-1}\gradthree{i}{t}{k}}}_{\mathcal{A}_1}
\end{align}
We first find a upper bound for $\mathcal{A}_1$ in accordance with Lemma \ref{lemma:1}:
\begin{align}
    \mathcal{A}_1 &= 2\eta\sum_{i=1}^M\sum_{k=0}^{K_i-1}\w_i\cdot\E\innerproduct{\optxs - \mathbf{x}_t}{\gradthree{i}{t}{k}} \nonumber \\
    % &\leq 2\eta\sum_{i=1}^M\sum_{k=0}^{K_i-1}\w_i\cdot\E\left[F_i\left(\optxs\right) - F_i\left(\mathbf{x}_t\right)+L\norm{\lx{i}{t}{k}-\mathbf{x}_t}-\frac{\mu}{4}\norm{\mathbf{x}_t - \optxs} \right] \\
    &\leq 2\eta\sum_{i=1}^M\w_iK_i\left[F_i\left(\optxs\right) - F_i\left(\mathbf{x}_t\right)\right] - \frac{\eta\mu\bar{K}}{2}\norm{\mathbf{x}_t - \optxs} \\
    &\quad +{2\eta L}\underbrace{\sum_{i=1}^M\sum_{k=0}^{K_i-1}\w_i\E\norm{\lx{i}{t}{k}-\mathbf{x}_t}}_{\mathcal{A}_3}
\end{align}
To find the maximum value for $\mathcal{A}_3$, we bound $\E\norm{\lx{i}{t}{k}-\mathbf{x}_t}$ for $k \in \{1, ..., K_i\}$ via the following inequality: 
\begin{align*}
    &\quad \E\norm{\lx{i}{t}{k}-\mathbf{x}_t}\\
    &= \E\norm{\lx{i}{t}{k-1}-\eta \nabla f_i\left(\lx{i}{t}{k-1},\varepsilon_{k-1}^{(i)}\right)-\mathbf{x}_t} \\
    &\leq \E\norm{\lx{i}{t}{k-1}-\mathbf{x}_t-\eta\gradthree{i}{t}{k-1}} + \eta^2\sigma^2 \\
    &\overset{(a)}{\leq} \left(1+\frac{1}{K_i-1}\right)\norm{\lx{i}{t}{k-1}-\mathbf{x}_t} +\eta^2\sigma^2 \\
    & \quad + K_i\eta^2\bracket{2\norm{\gradthree{i}{t}{k-1} - \grad{i}{t}} + 2\norm{\grad{i}{t}}} \\
    &\leq \left(1+\frac{1}{K_i-1}+{2K_i\eta^2L^2}\right)\norm{\lx{i}{t}{k-1}-\mathbf{x}_t}\\
    &\quad +{\eta^2\sigma^2}+2K_i\eta^2\norm{\gradtwos{i}{t}} 
\end{align*}
where $(a)$ follows triangle inequality, i.e., $(x + y)^2 \leq (1+k)x^2 + (1+1/k)y^2$ for all $k>0$. By setting $\eta \leq \sqrt{\frac{1}{2K_{\max}\left(K_{\max}-1\right)L^2}}$, we have: 
\begin{align}
    &\quad \E\norm{\lx{i}{t}{k}-\mathbf{x}_t} \nonumber \\
    &\leq \sum_{\kappa=0}^{k-1}\left(1+\frac{2}{K_i-1}\right)^\kappa \left(\eta^2\sigma^2+{2K_i\eta^2}\norm{\gradtwos{i}{t}}\right) \nonumber \\
    &= \frac{\left(1+\frac{2}{K_i-1}\right)^{k}-1}{\frac{2}{K_i-1}}\left({\eta^2\sigma^2}+{2K_i\eta^2}\norm{\gradtwos{i}{t}}\right) \nonumber \\
    % &\leq \frac{K_i-1}{2}\cdot\left[\left(1+\frac{2}{K_i-1}\right)^{K_i}-1\right]\left({\eta^2\sigma^2}+{2K_i\eta^2}\norm{\gradtwos{i}{t}}\right) \\
    &\overset{(a)}{\leq} 4K_i\left(\eta^2\sigma^2+{2K_i\eta^2}\norm{\gradtwos{i}{t}}\right) \label{eq:1-1}
\end{align}
where $(a)$ is on account for:
\begin{equation*}
    \left(1+\frac{2}{K_i-1}\right)^{K_i} \leq 9\qquad \text{for}\qquad K_i \geq 2
\end{equation*}
Based on the derivative above, we can obtain the bound for $\mathcal{A}_3$ with:
\begin{align*}
    \mathcal{A}_3 &\leq \sum_{i=1}^M\sum_{k=0}^{K_i-1}\w_i\cdot4K_i\left({\eta^2\sigma^2}+{2K_i\eta^2}\norm{\gradtwos{i}{t}}\right) \\
    &={4\eta^2\sigma^2}\sum_{i=1}^M\w_iK_i^2+{8\eta^2}\sum_{i=1}^M\w_iK_i^3\norm{\gradtwos{i}{t}}
\end{align*}
Therefore, the bound for $\mathcal{A}_1$ can be further simplified as:
\begin{align*}
    \mathcal{A}_1 &\leq {2\eta}\sum_{i=1}^{M}\w_iK_i\left[F_i\left(\optxs\right) - F_i\left(\mathbf{x}_t\right)\right]-\frac{\eta\mu\bar{K}}{2}\norm{\mathbf{x_t}-\optx}\\
    & \quad+ {2\eta L} \left[{4\eta^2\sigma^2}\sum_{i=1}^M \w_i K_i^2 + {8\eta^2} \sum_{i=1}^M \w_i K_i^3 \norm{\gradtwos{i}{t}}\right]
\end{align*}
Next, we consider the bound for $\mathcal{A}_2$:
\begin{align*}
    \mathcal{A}_2 &\leq {2\eta^2}\cdot\E\norm{\sum_{i=1}^M\sum_{k=0}^{K_i-1}\w_i\left[\gradthree{i}{t}{k}-\gradtwos{i}{t}\right]}\\
    &\quad + {2\eta^2}\cdot\E\norm{\sum_{i=1}^M\w_iK_i\gradtwos{i}{t}} \\
    &\leq {2\eta^2L^2}\sum_{i=1}^M\sum_{k=0}^{K_i-1}\w_iK_i\cdot\E\norm{\lx{i}{t}{k}-\mathbf{x}_t}\\
    &\quad+{2\eta^2}\sum_{i=1}^M\w_iK_i^2\norm{\gradtwo{i}{t}} \\
    % &\leq {2\eta^2L^2}\sum_{i=1}^M\sum_{k=0}^{K_i-1}\w_iK_i\cdot\left[4K_i\left({\eta^2\sigma^2}+{2\eta^2K_i}\norm{\gradtwo{i}{t}}\right)\right]+{2\eta^2}\sum_{i=1}^M\w_iK_i^2\norm{\gradtwo{i}{t}} \\
    &\leq {8\eta^4L^2\sigma^2}\sum_{i=1}^M\w_iK_i^3\\
    &\quad +{2\eta^2}\sum_{i=1}^M\w_iK_i^2\left(1+{8\eta^2L^2K_i^2}\right)\norm{\gradtwo{i}{t}}
\end{align*}
Therefore, pluging $\mathcal{A}_1$ and $\mathcal{A}_2$ into the inequality bound for $\E\norm{\mathbf{x}_{t+1}-\optxs}$, the bound can be simplified as:
\begin{align*}
    &\quad \E\norm{\mathbf{x}_{t+1}-\optxs}\\
    % &\leq \left(1-\frac{\eta\mu\bar{K}}{2} + {16\eta^3L^3K_{\max}^3B^2} + {2\eta^2K_{\max}^2B^2} \right)\E\norm{\mathbf{x}_{t}-\optxs} + {2\eta}\sum_{i=1}^M\w_iK_i\left[F_i(\optxs)-F_i\left(\mathbf{x}_t\right)\right]\\
    % &\qquad + {8\eta^3\sigma^2L}\sum_{i=1}^M\w_iK_i^2 + {2\eta}\sum_{i=1}^M\w_i^2K_i + {8\eta^4\sigma^2L^2}\sum_{i=1}^M\w_iK_i^3\\
    &\leq \left(1-\frac{\eta\mu\bar{K}}{2}\right)\E\norm{\mathbf{x}_t-\optxs}\\
    &\quad +2\eta\sum_{i=1}^M\w_iK_i\left[F_i(\optxs)-F_i\left(\mathbf{x}_t\right)\right]\\
    &\quad +8\eta^3\sigma^2L\sum_{i=1}^M\w_iK_i^2+\eta^2\sigma^2\sum_{i=1}^M\w_i^2K_i\\
    &\quad +8\eta^4L^2\sigma^2\sum_{i=1}^M\w_iK_i^3+16\eta^3L\sum_{i=1}^M\w_iK_i^3\norm{\gradtwos{i}{t}}\\
    &\quad +2\eta^2\sum_{i=1}^M\w_iK_i^2\left(1+8\eta^2L^2K_i\right)\norm{\gradtwos{i}{t}}
\end{align*}
Divided $\bar{K}$ on the both side, we can obtain the following formula: 
\begin{align*}
    &\quad \frac{1}{\bar{K}}\sum_{i=1}^M\w_iK_i\left[F_i(\optxs)-F_i\left(\mathbf{x}_t\right)\right] \\
    &\leq \frac{1}{\bar{K}\eta}\left(1-\frac{\eta\mu\bar{K}}{2}\right)\E\norm{\mathbf{x}_t-\optxs}-\frac{1}{\bar{K}\eta}\E\norm{\mathbf{x}_{t+1}-\optxs} \\
    & \quad + \frac{8\eta^2\sigma^2L}{\bar{K}}\sum_{i=1}^M\w_iK_i^2+\frac{\eta\sigma^2}{\bar{K}}\sum_{i=1}^M\w_i^2K_i\\
    &\quad+\frac{8\eta^3L^2\sigma^2}{\bar{K}}\sum_{i=1}^M\w_iK_i^3
\end{align*}
By applying Lemma 1 from \cite{karimireddy2020scaffold}, we can obtain the desirable result. It is worthwhile to mention a formula below that supports the reason why the gap exists between a stable point and the optimal solution: 
\begin{align*}
    &\quad \sum_{i=1}^M \w_i K_i (F_i(\mathbf{x}_t) - F_i(\x))\\
    &\geq K_{\min} \sum_{i=1}^M \w_i (F_i(\mathbf{x}_t) - F_i(\x))\\
    &\quad - \sum_{i=1}^M \w_i (K_i-K_{\min})F_i(\x) 
\end{align*}
The inequality holds when the value of the objective function is non-negative. This formula indicates that the data heterogeneity can be eliminated under homogeneous computing environment since for all $i \in \{1, ..., M\}$, $K_i = K_{\min}$. Thus, in this case, we can obtain the same convergence order as \cite{karimireddy2020scaffold}. 

% \section{Preliminary for Algorithm \ref{algo:1}}

\section{Proof of Theorem \ref{tm:tm2}} \label{proof:tm2}

According to $L$-smooth, we have: 
\begin{align}
    &\quad \E\left[F(\gx{t+1})\right] - F(\gx{t}) \nonumber \\
    &\leq \E\innerproduct{\grad{t}}{\gx{t+1}-\gx{t}} + \frac{L}{2} \E \norm{\gx{t+1}-\gx{t}} \label{eq:2-1}
\end{align}
\textbf{The first term of Equation (\ref{eq:2-1}).} We firstly find the bound for the first term of Equation (\ref{eq:2-1}):
\begin{small}
\begin{align}
    &\quad \E\innerproduct{\grad{t}}{\gx{t+1}-\gx{t}}\nonumber \\
    &= \E\left\langle\grad{t}, -\eta \sum_{i=1}^M\sum_{k=0}^{K_i-1}\w_i\grad{i}{t}{k} \right.\nonumber \\
    % &\qquad \nonumber \\
    &\qquad +\eta\lambda\sum_{i=1}^M \w_i \sum_{\kappa=0}^{K_i-1}\grad{i}{t-1}{\kappa} -\eta\lambda\bar{K} \grad{t-1} \nonumber \\
    &\qquad \left. -\eta\lambda\bar{K} \sum_{i,K_i \leq \bar{K}} \sum_{k=0}^{K_i-1} \frac{\w_i}{K_i} \bracket{\grad{i}{t-1}{k}-\grad{i}{t-1}} \right\rangle \nonumber \\
    % &\qquad \left. -\eta\lambda\bar{K} \grad{t-1} \right\rangle \nonumber \\
    % = &-\tilde{\eta}\E\innerproduct{\grad{t}}{\sum_{i=2}^M\sum_{k=0}^{K_i-1}\w_i\grad{i}{t}{k}-\lambda\sum_{i=1}^M \w_i \sum_{\kappa=0}^{K_i-1}\grad{i}{t-1}{\kappa}+\lambda\bar{K} \sum_{i,K_i \leq \bar{K}} \frac{\w_i}{K_i} \sum_{k=0}^{K_i-1} \bracket{\grad{i}{t-1}{k}-\grad{i}{t-1}} + \lambda\bar{K} \grad{t-1} } \\
    &= -\frac{\eta\lambda}{\bar{K}}\E\left\langle\bar{K}\grad{t}, \sum_{i=1}^M\sum_{k=0}^{K_i-1}\w_i\grad{i}{t}{k}\right. \nonumber \\
    &\qquad \qquad \quad -\sum_{i=1}^M  \sum_{\kappa=0}^{K_i-1} \w_i \grad{i}{t-1}{\kappa} + \bar{K} \grad{t-1} \nonumber \\
    &\qquad \qquad \quad \left. +\bar{K} \sum_{i,K_i \leq \bar{K}} \sum_{k=0}^{K_i-1} \frac{\w_i}{K_i} \bracket{\grad{i}{t-1}{k}-\grad{i}{t-1}} \right\rangle \nonumber \\
    &\quad -\eta (1-\lambda) K_{\max} \E\innerproduct{\grad{t}}{\frac{1}{K_{\max}}\sum_{i=1}^M\sum_{k=0}^{K_i-1}\w_i\grad{i}{t}{k}} \nonumber \\
    % = &-\frac{\eta\lambda}{\bar{K}}\E\innerproduct{\bar{K}\grad{t}}{\sum_{i=2}^M\sum_{k=0}^{K_i-1}\w_i\grad{i}{t}{k}-\sum_{i=1}^M \w_i \sum_{\kappa=0}^{K_i-1}\grad{i}{t-1}{\kappa}+\bar{K} \sum_{i,K_i \leq \bar{K}} \frac{\w_i}{K_i} \sum_{k=0}^{K_i-1} \bracket{\grad{i}{t-1}{k}-\grad{i}{t-1}} + \bar{K} \grad{t-1} } \\
    % &-\eta K_{max}\E\innerproduct{\grad{t}}{\frac{1-\lambda}{K_{max}}\sum_{i=2}^M\sum_{k=0}^{K_i-1}\w_i\grad{i}{t}{k}} \\
    &=-\frac{\eta\lambda\bar{K}}{2} \norm{\grad{t}} \label{eq:2-2} \\
    &\quad -\frac{\eta\lambda}{2\bar{K}}
    \E\left\|\sum_{i=1}^M \sum_{k=0}^{K_i-1}\w_i\grad{i}{t}{k} \right. \nonumber\\
    &\qquad \qquad \quad -\sum_{i=1}^M \sum_{\kappa=0}^{K_i-1}\w_i\grad{i}{t-1}{\kappa} + \bar{K}\grad{t-1} \nonumber\\
    &\qquad \qquad \quad \left. +\bar{K}\sum_{i,K_i \leq \bar{K}}\sum_{k=0}^{K_i-1} \frac{\w_i}{K_i} \bracket{\grad{i}{t-1}{k}-\grad{i}{t-1}} \right\|_2^2 \label{eq:2-3}\\
    &\quad +\frac{\eta\lambda}{2\bar{K}}\E\left\|\bar{K}\grad{t}-\bar{K}\grad{t-1} \right. \nonumber\\
    &\qquad \qquad \quad -\sum_{i=1}^M \sum_{k=0}^{K_i-1}\w_i\left(\grad{i}{t}{k}-\grad{i}{t-1}{k}\right) \nonumber \\ 
    &\qquad \qquad \quad \left. -\bar{K}\sum_{i,K_i\leq\bar{K}}\sum_{k=0}^{K_i-1} \frac{\w_i}{K_i} \left(\grad{i}{t-1}{k}-\grad{i}{t-1}\right)\right\|_2^2 \label{eq:2-4}\\
    & \quad -\frac{\eta(1-\lambda) K_{\max}}{2}\E\norm{\grad{t}} \label{eq:2-5}\\ 
    &\quad -\frac{\eta(1-\lambda)}{2 K_{\max}}\E\norm{\sum_{i=1}^M \sum_{k=0}^{K_i-1}\w_i\grad{i}{t}{k}} \label{eq:2-6}\\
    &\quad +\frac{\eta(1-\lambda) K_{\max}}{2}\E\left\|\grad{t}\right. \nonumber\\
    &\qquad \qquad \qquad \qquad \qquad \left.-\frac{1}{K_{\max}}\sum_{i=1}^M \sum_{k=0}^{K_i-1}\w_i\grad{i}{t}{k}\right\|_2^2 \label{eq:2-7}
\end{align}
\end{small}
Next, we bound the term of Equation (\ref{eq:2-7}) ignoring the coefficient term, i.e., $\frac{\eta(1-\lambda) K_{\max}}{2}$: 
\begin{align*}
    &\quad\E\left\|\grad{t} - \frac{1}{K_{\max}}\sum_{i=1}^M \sum_{k=0}^{K_i-1}\w_i\grad{i}{t}{k}\right\|_2^2 \\
    &\leq \E\left\|\sum_{i=1}^M\w_i\bracket{1-\frac{K_i}{K_{\max}}}\grad{t}\right.\\
    &\qquad \left.-\frac{1}{K_{\max}}\sum_{i=1}^M \sum_{k=0}^{K_i-1}\w_i\left(\grad{i}{t}-\grad{i}{t}{k}\right)\right\|_2^2 \\
    &\overset{(a)}{\leq} \bracket{1+\frac{K_{\min}}{K_{\max}}}\E\norm{\sum_{i=1}^M\w_i\bracket{1-\frac{K_i}{K_{\max}}}\grad{i}{t}} \\
    % &\quad +\bracket{1+\frac{K_{\max}}{K_{\min}}}\E\norm{\frac{1}{K_{\max}}\sum_{i=1}^M \sum_{k=0}^{K_i-1}\w_i\left[\grad{i}-\grad{i}{t}{k}\right]} \\
    &\quad +\bracket{1+\frac{K_{\max}}{K_{\min}}} T_1\\
    &\leq\bracket{1-\frac{\bar{K}}{K_{\max}}} B^2\E\norm{\grad{t}} +\bracket{1+\frac{K_{\max}}{K_{\min}}} T_1 
    % &\quad+\bracket{1+\frac{K_{\max}}{K_{\min}}}\E\norm{\frac{1}{K_{\max}}\sum_{i=1}^M \sum_{k=0}^{K_i-1}\w_i\left[\grad{i}-\grad{i}{t}{k}\right]}
\end{align*}
where 
\begin{align*}
    T_1 &= \E\norm{\frac{1}{K_{\max}}\sum_{i=1}^M \sum_{k=0}^{K_i-1}\w_i\left(\grad{i}{t}-\grad{i}{t}{k}\right)}\\
    &\leq \frac{L^2}{K_{\max}^2} \sum_{i=1}^M \sum_{k=0}^{K_i-1} \w_i K_i \E\norm{\x{i}{t}{k} - \x{t}}
\end{align*}
and $(a)$ follows triangle inequality, i.e., $(x + y)^2 \leq (1+c)x^2 + (1+1/c)y^2$ for all $c>0$. We denote Equation (\ref{eq:2-4}) omitted the coefficient $\frac{\eta \lambda}{2 \bar{K}}$ by $T_2$. Therefore, its upper bound is obtained through the following derivation: 
\begin{small}
\begin{align}
    T_2 &\overset{(a)}{\leq} 5\bar{K}^2 \E\norm{\grad{t}-\grad{t-1}} \nonumber\\
    &\quad+5\E\norm{\sum_{i=1}^M \sum_{k=0}^{K_i-1}\w_i\left(\grad{i}{t}{k}-\grad{t}\right)} \nonumber\\
    &\quad +5\E\norm{\sum_{i=1}^M \sum_{k=0}^{K_i-1}\w_i\left(\grad{i}{t-1}{k}-\grad{t-1}\right)} \nonumber\\
    &\quad +5\E\norm{\sum_{i=1}^M \sum_{k=0}^{K_i-1}\w_i\left(\grad{t}-\grad{t-1}\right)} \nonumber\\
    &\quad +5\bar{K}^2\E\norm{\sum_{i,K_i \leq \bar{K}}\sum_{k=0}^{K_i-1} \frac{\w_i}{K_i} \left(\grad{i}{t-1}{k}-\grad{i}{t-1}\right)} \nonumber\\
    &\leq 5 L^2 \bracket{\bar{K}^2 + \sum_{i=1}^M \w_i K_i^2} \E\norm{\gx{t}-\gx{t-1}} \nonumber\\
    &\quad + 5L^2 \sum_{i=1}^M \sum_{k=0}^{K_i-1} \w_i K_i \E\norm{\lx{i}{t}{k}-\gx{t}} \nonumber\\
    &\quad+ 5L^2 \sum_{i=1}^M \sum_{k=0}^{K_i-1} \w_i K_i \E\norm{\lx{i}{t-1}{k}-\gx{t-1}} \nonumber\\
    % &\quad+5L^2 \sum_{i=1}^M \w_i K_i^2 \E\norm{\gx{t}-\gx{t-1}} \nonumber\\
    &\quad +5\bar{K}^2 L^2\sum_{i,K_i \leq \bar{K}} \sum_{k=0}^{K_i-1} \frac{\w_i}{K_i} \E\norm{\lx{i}{t-1}{k}-\gx{t-1}} \nonumber \\
    &\leq 10 L^2 \sum_{i=1}^M \w_i K_i^2 \E\norm{\gx{t}-\gx{t-1}} \nonumber\\
    &\quad + 5L^2 \sum_{i=1}^M \sum_{k=0}^{K_i-1} \w_i K_i \E\norm{\lx{i}{t}{k}-\gx{t}} \nonumber\\
    &\quad+ 5L^2 \sum_{i=1}^M \sum_{k=0}^{K_i-1} \w_i \bracket{K_i + \frac{\bar{K}^2}{K_i}} \E\norm{\lx{i}{t-1}{k}-\gx{t-1}} \label{eq:2-8}
    % &\quad +5\bar{K}^2 L^2\sum_{i,K_i \leq \bar{K}} \sum_{k=0}^{K_i-1} \frac{\w_i}{K_i} \E\norm{\lx{i}{t-1}{k}-\gx{t-1}} \label{eq:2-8}
\end{align}
\end{small}
\noindent where $(a)$ divides $\left(\grad{i}{t}{k}-\grad{i}{t-1}{k}\right)$ into three terms, i.e., (each bracket should be treated as an individual term): $\left(\grad{i}{t}{k}-\grad{i}{t}\right) - \left(\grad{i}{t-1}{k}-\grad{i}{t-1}\right) + \left(\grad{i}{t}-\grad{i}{t-1}\right)$. By observing Equation (\ref{eq:2-8}), we notice that it is indispensable to acquire the upper limit of $\E\norm{\lx{i}{t}{k}-\gx{t}}$: 
\begin{small}
\begin{align}
    &\quad\E\norm{\lx{i}{t}{k}-\gx{t}} \nonumber \\
    &= \E \norm{\lx{i}{t}{k-1}-\eta\left[g_{t,k-1}^{(i)}+\lambda\left(\nu-\nu^{(i)}\right)\right]-\gx{t}} \nonumber \\
    &\leq \bracket{1+\frac{1}{K_i-1}} \E\norm{\lx{i}{t}{k-1}-\gx{t}} \nonumber \\
    &\quad + K_i \eta^2 \E \left\|g_{t,k-1}^{(i)}-\frac{\lambda}{K_i}\sum_{\kappa=0}^{K_i-1}g_{t-1,\kappa}^{(i)}+\lambda \sum_{j=1}^M \sum_{k=0}^{K_j-1} \frac{\w_j}{K_j} g_{t-1,k}^{(j)} \right. \nonumber \\
    &\qquad \qquad \qquad \left.+ \lambda\sum_{j,K_j > \bar{K}} \sum_{k=0}^{K_j-1} \frac{\w_j}{K_j} \bracket{g_{t-1,0}^{(j)}-g_{t-1,k}^{(j)}}\right\|_2^2 \nonumber \\
    &\leq \bracket{1+\frac{1}{K_i-1}} \E\norm{\lx{i}{t}{k-1}-\gx{t}} \label{eq:2-9} \\
    &\quad + K_i\eta^2 \E\left\|\grad{i}{t}{k-1} - \frac{\lambda}{K_i} \sum_{\kappa=0}^{K_i-1}\grad{i}{t-1}{\kappa}\right.\nonumber \\
    & \qquad \qquad \qquad + \lambda\sum_{j=1}^M \frac{\w_j}{K_j}\sum_{k=0}^{K_j-1} \grad{j}{t-1}{k} \nonumber\\
    &\qquad \qquad \qquad \left. +\lambda\sum_{j,K_j > \bar{K}}\frac{\w_j}{K_j} \sum_{k=0}^{K_j-1}\left(\grad{j}{t-1}-\grad{j}{t-1}{k}\right)\right\|_2^2 \label{eq:2-10}\\
    &\quad + 4K_i^2\eta^2\sigma^2 + 4\lambda^2\eta^2\sigma^2 + 12 K_i^2 \eta^2 \lambda^2\sigma^2\sum_{j=1}^M\frac{\w_j^2}{K_j} \label{eq:2-11}
\end{align}
\end{small}
We denote the second norm in Equation (\ref{eq:2-10}) by $T_3$. Similar to the derivation for $T_2$, i.e., Equation (\ref{eq:2-8}), we have the following inequality under Assumption \ref{ass:4}: 
\begin{align}
    T_3 \leq & 8L^2\E\norm{\lx{i}{t}{k-1}-\gx{t}} + \frac{8\lambda^2L^2}{K_i} \sum_{\kappa=0}^{K_i-1}\E\norm{\lx{i}{t-1}{\kappa}-\gx{t-1}} \nonumber \\
    &+16\lambda^2L^2 \sum_{j=1}^M \sum_{k=0}^{K_j-1} \frac{\w_j}{K_j} \E\norm{\lx{j}{t-1}{k}-\gx{t-1}} \nonumber \\
    &+ 8 \bracket{(1-\lambda)^2 B^2 + \lambda^2} \E\norm{\grad{t}} \nonumber \\
    &+ 16\lambda^2L^2 \E\norm{\gx{t}-\gx{t-1}} %\label{eq:2-12}
\end{align}
By setting $\eta \leq \frac{1}{2\sqrt{2}L K_{\max}}$ and following the steps of Equation (\ref{eq:1-1}), we have: 
\begin{align}
    \E\norm{\lx{i}{t}{k}-\gx{t}} \leq \frac{K_i-1}{2}\left(\bracket{1+\frac{2}{K_i-1}}^{k+1}-1\right)T_4
\end{align}
where
\begin{align*}
    T_4 &=16\lambda^2L^2K_i\eta^2 \norm{\gx{t}-\gx{t-1}}\\
    &\quad +8\lambda^2L^2\eta^2\sum_{k=0}^{K_i-1}\norm{\lx{i}{t-1}{k}-\gx{t-1}} \\
    &\quad +16\lambda^2L^2K_i\eta^2\sum_{j=1}^M \sum_{k=0}^{K_j-1} \frac{\w_j}{K_j} \norm{\lx{j}{t-1}{k}-\gx{t-1}} \\
    &\quad + 8 K_i \eta^2 \bracket{(1-\lambda)^2 B^2 + \lambda^2} \norm{\grad{t}} +4K_i\eta^2\sigma^2 \\
    &\quad + 4 \lambda^2\eta^2\sigma^2 + 12K_i\eta^2\lambda^2\sigma^2\sum_{j=1}^M \frac{\w_j^2}{K_j}. 
\end{align*}
As a result, when the learning rate $\eta$ is sufficiently small, the upper bound for $T_2$ should be 
\begin{align}
    T_2 \leq&\bracket{20L^2 \sum_{i=1}^M \w_i K_i^2}\E\norm{\gx{t}-\gx{t-1}} \nonumber\\
    &+20\bar{K}^2L^2  \sum_{i=1}^M \sum_{k=0}^{K_i-1}\w_i \E\norm{\lx{i}{t-1}{k}-\gx{t-1}} \nonumber\\
    % &+\left[ 60\eta^2L^2(1-\lambda)^2K_{max}^4 B^2+60\eta^2L^2\lambda^2 \sum_{i=1}^M \w_i K_i^4 \right]  \norm{\grad{t}} \\
    &+60\eta^2L^2 K_{\max}^4\left((1-\lambda)^2 B^2+\lambda^2\right) \E\norm{\grad{t}} \nonumber\\
    &+30 \eta^2 L^2 \sigma^2 \sum_{i=1}^M \w_i K_i^4 + 30 \eta^2 L^2 \sigma^2 \lambda^2\sum_{i=1}^M \w_i K_i^3 \nonumber\\
    &+ 90 \eta^2 L^2 \sigma^2\lambda^2 \bracket{\sum_{j=1}^M\frac{\w_j^2}{K_j}}\sum_{i=1}^M \w_i K_i^4 \label{eq:2-12}
\end{align}
\textbf{The second term of Equation (\ref{eq:2-1}).} We now give the upper limit for $\E \norm{\gx{t+1}-\gx{t}}$: 
\begin{align}
    &\quad \E \norm{\gx{t+1}-\gx{t}} \nonumber \\
    &\leq \eta^2 \E\left\|\sum_{i=1}^M \sum_{k=0}^{K_i-1}\w_i \grad{i}{t}{k} \right. \nonumber \\
    &\qquad \quad + \lambda \sum_{i,K_i > \bar{K}}\sum_{k=0}^{K_i-1}\frac{\w_i\bar{K}}{K_i}\bracket{\grad{i}{t-1}-\grad{i}{t-1}{k}} \nonumber \\
    &\qquad \quad +\left.\lambda \sum_{i=1}^M \sum_{k=0}^{K_i-1}\w_i\bracket{\frac{\bar{K}}{K_i}-1}\grad{i}{t-1}{k} \right\|_2^2 \nonumber \\
    &\leq 2\eta^2\lambda^2\E\left\|\sum_{i=1}^M \sum_{k=0}^{K_i-1}\w_i\grad{i}{t}{k}-\sum_{i=1}^M \sum_{\kappa=0}^{K_i-1}\w_i\grad{i}{t-1}{\kappa}
    \right. \nonumber \\
    &\quad \quad \quad \quad +\sum_{i,K_i \leq \bar{K}}\frac{\w_i\bar{K}}{K_i}\sum_{k=0}^{K_i-1}\bracket{\grad{i}{t-1}{k}-\grad{i}{t-1}} \nonumber \\
    &\quad \quad \quad \quad \left.+\bar{K}\grad{t-1}\right\|_2^2  \nonumber \\
    &\quad+2\eta^2(1-\lambda)^2\norm{\sum_{i=1}^M \sum_{k=0}^{K_i-1}\w_i \grad{i}{t}{k}} \nonumber \\
    &\quad+3\eta^2\sigma^2\sum_{i=1}^M\w_i^2K_i + 3\eta^2\lambda^2\sigma^2\sum_{i=1}^M\w_i^2\bracket{\frac{5\bar{K}^2}{K_i}+K_i} \label{eq:2-13}
    % &\quad+12\eta^2\lambda^2\sigma^2\sum_{i,K_i > \bar{K}} \frac{\w_i^2\bar{K}^2}{K_i} \label{eq:2-13}
\end{align}
\textbf{Final result.} By the inequality from Equation (\ref{eq:2-12}) and Equation (\ref{eq:2-13}), we can add two extra terms on the left-hand side of Equation (\ref{eq:2-1}), i.e., $\E\norm{\x{t} - \x{t-1}}$ and $\sum_{i=1}^M \sum_{k=0}^{K_i-1}\w_i K_i \E\norm{\x{i}{t-1}{k} - \x{t-1}}$, and obtain the following bound when the learning rate is sufficiently small: 
\begin{footnotesize}
\begin{align*}
    &\quad \E[F(\gx{t+1})] + p_1 \E\norm{\gx{t+1}-\gx{t}} +p_2 \sum_{i=1}^M \sum_{k=0}^{K_i-1}\w_i K_i \E\norm{\lx{i}{t}{k}-\gx{t}} \\
    &\leq F(\gx{t}) + p_1 \E\norm{\gx{t}-\gx{t-1}} + p_2 \sum_{i=1}^M \sum_{k=0}^{K_i-1}\w_i K_i \E\norm{\lx{i}{t-1}{k}-\gx{t-1}} \\
    & \quad -\left(\frac{\eta\lambda\bar{K}}{2}+\frac{\eta\bracket{1-\lambda}K_{\max}}{2}\bracket{1-\bracket{1-\frac{\bar{K}}{K_{\max}}}B^2}\right)\norm{\grad{t}} \\
    % &\quad +\left(\frac{\eta(1-\lambda)L^2}{K_{\min}} + \frac{5\eta \lambda L^2}{2\bar{K}} + p_2\right)\sum_{i=1}^M \sum_{k=0}^{K_i-1}\w_i K_i \E\norm{\lx{i}{t}{k}-\gx{t}} \\
    &\quad + 3\bracket{\frac{L}{2}+p_1} \eta^2 \sigma^2 \left(\sum_{i=1}^M \w_i^2 K_i +\lambda^2 \sum_{i=1}^M\w_i^2\bracket{\frac{5\bar{K}^2}{K_i}+K_i}\right)
\end{align*}
\end{footnotesize}
where $p_1 = o\bracket{\frac{\eta\lambda}{\bar{K}} L^2 \sum_{i=1}^M \w_i \left(K_i^2+\bar{K}^2\right)}$ and $p_2 = o\bracket{\frac{\eta\lambda}{\bar{K}} L^2 \bracket{1+\frac{\bar{K}^2}{K_{\min}^2}}}$. Therefore, the final result is:
\begin{footnotesize}
\begin{align*}
    &\quad\frac{1}{T} \sum_{t=1}^T \norm{\grad{t}} \\
    % \leq&\mathcal{O} \bracket{\frac{F(\optx)-F(\gx{1})}{\eta\lambda \bar{K}T}} +\mathcal{O} \bracket{\frac{\eta\sigma^2 L}{\lambda \bar{K}}\left[\sum_{i=1}^M \w_i^2 K_i+\lambda^2\sum_{i=1}^M \w_i^2 \bracket{\frac{\bar{K}}{K_i}-1}^2 K_i+\lambda^2\sum_{i,K_i> \bar{K}}\frac{\w_i^2 \bar{K}}{K_i}\right]} \\
    &= \mathcal{O} \bracket{\frac{F(\optx)-F(\gx{1})}{\eta\lambda \bar{K}T}} + \mathcal{O} \bracket{\frac{\eta\sigma^2 L}{\lambda \bar{K}}\sum_{i=1}^M \w_i^2 K_i}\\
    &\quad +  \mathcal{O} \bracket{\frac{\eta\sigma^2 L\lambda}{ \bar{K}}\sum_{i=1}^M \w_i^2 \bracket{\frac{\bar{K}}{K_i}-1}^2} +  \mathcal{O} \bracket{\eta\sigma^2 L\lambda\sum_{i,K_i> \bar{K}}\frac{\w_i^2 \bar{K}}{K_i}}
\end{align*}
\end{footnotesize}

\section{Proof of Theorem \ref{tm:tm3}} \label{proof:tm3}

\newcommand{\Q}{\mathcal{Q}}

At the very beginning, we set $\lambda = 1$ to find a valid bound. Based on the definition, we can find a recursion function for $\E\norm{\gx{t+1}-\optx}$: 
\begin{align}
    &\quad \E\norm{\gx{t+1}-\optx} = \E\norm{\bracket{\gx{t} - \optx} + \bracket{\gx{t+1} - \optx}} \nonumber \\
    &= \E\norm{\gx{t} - \optx} + \E\norm{\gx{t+1} - \gx{t}} + 2\E\innerproduct{\x{t}-\x}{\x{t+1}-\x{t}} \nonumber\\
    &= \E\norm{\gx{t} - \optx} + \E\norm{\gx{t+1} - \gx{t}} \label{eq:3-1}\\
    % &\quad + 2\underbrace{\E\innerproduct{\x{t} - \x}{-\eta \sum_{i=1}^M \sum_{k=0}^{K_i-1}\w_i\grad{i}{t}{k}}}_{\mathcal{Q}_1} \label{eq:3-2}\\
    &\quad + 2\E\innerproduct{\x{t} - \x}{-\eta \sum_{i=1}^M \sum_{k=0}^{K_i-1}\w_i\grad{i}{t}{k}} \label{eq:3-2}\\
    &\quad + 2\E\innerproduct{\x{t} - \x}{\eta \sum_{i=1}^M \sum_{k=0}^{K_i-1} \w_i\grad{i}{t-1}{k}} \label{eq:3-3} \\
    &\quad + 2\E\left\langle \x{t} - \x, \right. \nonumber \\
    &\quad \qquad \left. - \eta \bar{K} \sum_{i, K_i \leq \bar{K}} \frac{\w_i}{K_i} \sum_{k=0}^{K_i-1}\bracket{\grad{i}{t-1}{k} - \grad{i}{t-1}} \right\rangle \label{eq:3-4} \\
    % &\quad + 2{\underbrace{\E\innerproduct{\x{t} - \x}{- \eta \bar{K} \sum_{i, K_i \leq \bar{K}} \frac{\w_i}{K_i} \sum_{k=0}^{K_i-1}\bracket{\grad{i}{t-1}{k} - \grad{i}{t-1}}}}_{\mathcal{Q}_3}}\label{eq:3-4} \\
    &\quad + 2\E\innerproduct{\x{t} - \x}{-\eta \bar{K} \grad{t-1}} \label{eq:3-5}
\end{align}
We denote Equation (\ref{eq:3-2}) to (\ref{eq:3-5}) by $\Q_1$, $\Q_2$, $\Q_3$ and $\Q_4$, respectively. There are two terms in Equation (\ref{eq:3-2}) and Equation (\ref{eq:3-3}), namely $\mathcal{Q}_1$ and $\mathcal{Q}_2$, between which the subscript is different (i.e., one for $t$-th update while the others for $t-1$-th update). The following will present how to bound $\mathcal{Q}_1$ first, and then $\mathcal{Q}_2$. 
\begin{align*}
    \Q_1 &= \eta \sum_{i=1}^M \sum_{k=0}^{K_i-1}\w_i \E\innerproduct{\grad{i}{t}{k} - \grad{i}{t}}{\x - \x{t}}\\
    &\quad +  \eta \sum_{i=1}^M \w_i K_i \E\innerproduct{\grad{i}{t}}{\x - \x{t}} \\
    &\leq \frac{\eta}{2} \sum_{i=1}^M \sum_{k=0}^{K_i-1}\w_i \left(\frac{16}{\mu}\E\norm{\grad{i}{t}{k} - \grad{i}{t}} \right.\\
    &\qquad \qquad \qquad \qquad \left. + \frac{\mu}{16}\E\norm{\x - \x{t}}\right)\\
    &\quad +  \eta \sum_{i=1}^M \w_i K_i \E\innerproduct{\grad{i}{t}}{\x - \x{t}} \\
    &\leq \frac{8\eta L^2}{\mu} \sum_{i=1}^M \sum_{k=0}^{K_i-1}\w_i \E\norm{\x{i}{t}{k} - \x} + \frac{\eta\mu\bar{K}}{32} \E\norm{\x - \x{t}}\\
    &\quad + \eta \sum_{i=1}^M \w_i K_i \E\innerproduct{\grad{i}{t}}{\x - \x{t}}
\end{align*}
where the first inequality refers to $\langle a, b \rangle \leq (\|a\|_2^2 + \|b\|_2^2)/2$ and the last one is according to Assumption \ref{ass:1}. Likewise, we can find the bound for $\Q_2$: 
\begin{align*}
    \Q_2 &\leq \frac{8\eta L^2}{\mu} \sum_{i=1}^M \sum_{k=0}^{K_i-1}\w_i \E\norm{\x{i}{t-1}{k} - \x}\\
    &\quad + \frac{\eta\mu\bar{K}}{32} \E\norm{\x - \x{t}}\\
    &\quad +  \eta \sum_{i=1}^M \w_i K_i \E\innerproduct{\grad{i}{t-1}}{\x{t} - \x} \\
\end{align*}
As a result, we have: 
\begin{align*}
    &\quad \Q_1 + \Q_2\\
    &\leq \frac{8\eta L^2}{\mu} \sum_{i=1}^M \sum_{k=0}^{K_i-1}\w_i \bracket{\E\norm{\x{i}{t}{k} - \x} + \E\norm{\x{i}{t-1}{k} - \x}}\\
    &\quad + \eta \sum_{i=1}^M \w_i K_i \E\innerproduct{\grad{i}{t-1} - \grad{i}{t}}{\x{t} - \x} \\
    &\quad + \frac{\eta\mu\bar{K}}{16} \E\norm{\x - \x{t}} \\
    &\leq \frac{8\eta L^2}{\mu} \sum_{i=1}^M \sum_{k=0}^{K_i-1}\w_i \bracket{\E\norm{\x{i}{t}{k} - \x} + \E\norm{\x{i}{t-1}{k} - \x}}\\
    &\quad + \frac{3\eta\mu\bar{K}}{32} \E\norm{\x - \x{t}} + \frac{8\eta \bar{K} L^2}{\mu} \E\norm{\x{t} - \x{t-1}}
\end{align*}
where the last inequality is based on $\langle a, b \rangle \leq (\|a\|_2^2 + \|b\|_2^2)/2$ and Assumption \ref{ass:1}. 

For the term in Equation (\ref{eq:3-4}), according to the inequality that $\langle a, b \rangle \leq (\|a\|_2^2 + \|b\|_2^2)/2$ and the assumption of L-smooth, we have: 
\begin{align*}
    \Q_3 &= - \eta \bar{K} \sum_{i, K_i \leq \bar{K}} \frac{\w_i}{K_i} \sum_{k=0}^{K_i-1} \E\left\langle \x{t} - \x, \right. \\
    &\qquad \qquad \qquad \qquad \qquad \qquad \left. \grad{i}{t-1}{k} - \grad{i}{t-1} \right\rangle \\
    &\leq \frac{8\eta \bar{K} L^2}{\mu} \sum_{i, K_i \leq \bar{K}} \sum_{k=0}^{K_i-1} \frac{\w_i}{K_i} \E\norm{\x{i}{t-1}{k} - \x{t-1}}\\
    &\quad + \frac{\eta \mu \bar{K}}{32} \E\norm{\x - \x{t}}
\end{align*}
The last equality holds because $\w_i > 0$ and $K_i > 0$ such that the sum for those $K_i \leq \bar{K}$ is not greater than the one for all workers. 

% Prior to computing the bound for $\mathcal{Q}_4$,

Based on Lemma \ref{lemma:1} above, it is easy to derive the bound for $\mathcal{Q}_4$, which is: 
\begin{align*}
    &\quad \mathcal{Q}_4 = \eta \bar{K} \E\innerproduct{\grad{t-1}}{\x - \x{t}}\\
    &\leq \eta \bar{K} \bracket{F(\x) - F(\x{t}) - \frac{\mu}{4}\E\norm{\x{t}-\x} + L\E\norm{\x{t} - \x{t-1}}}
\end{align*}
According to the bound for $\E\norm{\x{t+1}-\x{t}}$ in the proof non-convex objectives, i.e., $T_2$, we have: 
\begin{align*}
    \E\norm{\x{t+1}-\x{t}} &\leq 2\eta^2\lambda^2 \mathcal{Q}_5 + 3\eta^2\sigma^2\sum_{i=1}^M \w_i^2 K_i\\
    &\quad + 3\eta^2 \lambda^2 \sigma^2 \sum_{i=1}^M \w_i^2 \bracket{\frac{\bar{K}}{K_i} - 1}^2 K_i\\
    &\quad + 12 \eta^2 \lambda^2 \sigma^2 \sum_{i, K_i > \bar{K}} \frac{\w_i^2 \bar{K}^2}{K_i}
\end{align*}
where 
\begin{align}
    \mathcal{Q}_5 &=  \E\left\|\sum_{i=1}^M \sum_{k=0}^{K_i-1}\w_i\grad{i}{t}{k} - \sum_{i=1}^M \sum_{k=0}^{K_i-1} \w_i\grad{i}{t-1}{k} \right.\nonumber\\
    &\qquad + \bar{K} \sum_{i, K_i \leq \bar{K}} \frac{\w_i}{K_i} \sum_{k=0}^{K_i-1}\bracket{\grad{i}{t-1}{k} - \grad{i}{t-1}}\nonumber\\
    &\qquad \left. + \bar{K} \grad{t-1} \right\|_2^2 \label{eq:3-6}
\end{align}
Unlike the procedure in non-convex objectives, $\mathcal{Q}_5$ cannot be eliminated. Therefore, we should find a general bound for Equation (\ref{eq:3-4}), where we can further simplify as: 
\begin{align*}
    \Q_5 &= \E\left\|\sum_{i=1}^M \sum_{k=0}^{K_i-1}\w_i\bracket{\grad{i}{t}{k} - \grad{i}{t}} \right. \\
    & \qquad - \sum_{i=1}^M \sum_{k=0}^{K_i-1} \w_i\bracket{\grad{i}{t-1}{k} - \grad{i}{t-1}}\\
    &\qquad + \bar{K} \sum_{i, K_i \leq \bar{K}} \frac{\w_i}{K_i} \sum_{k=0}^{K_i-1}\bracket{\grad{i}{t-1}{k} - \grad{i}{t-1}}\\
    &\qquad + \bar{K} \left(\grad{t-1} - \grad{t}\right)  + \bar{K}\grad{t}\\
    &\qquad \left.+ \sum_{i=1}^M \w_i K_i\bracket{\grad{i}{t} - \grad{i}{t-1}}\right\|_2^2\\
    &\leq 6L^2 \sum_{i=1}^M \sum_{k=0}^{K_i-1}\w_i K_i \E\norm{\x{i}{t}{k} - \x{t}} \\
    &\quad + 6L^2\sum_{i=1}^M \sum_{k=0}^{K_i-1}\w_i K_i \E\norm{\x{i}{t-1}{k} - \x{t-1}}\\
    &\quad + 6 \bar{K}^2 L^2 \sum_{i, K_i \leq \bar{K}} \sum_{k=0}^{K_i-1} \frac{\w_i}{K_i} \E\norm{\x{i}{t-1}{k} - \x{t-1}} \\
    &\quad + 12\bar{K}^2 L^2 \E\norm{\x{t} - \x{t-1}} + 6\bar{K}^2 L^2 \E\norm{\x{t} - \x}
\end{align*}
Plugging the results above, we can obtain the bound for $\E\|\x{t+1} - \x\|_2^2$ as: 
\begin{align*}
    &\qquad \E\norm{\gx{t+1}-\optx}\\
    &\leq \bracket{1-\frac{\eta\mu\bar{K}}{4}} \E\norm{\gx{t} - \optx} \\
    &\quad + 12\eta^2 \lambda^2 L^2 \sum_{i=1}^M \sum_{k=0}^{K_i-1}\w_i K_i \E\norm{\x{i}{t}{k} - \x{t}}\\
    &\quad + 12\eta^2 \lambda^2 L^2\sum_{i=1}^M \sum_{k=0}^{K_i-1}\w_i K_i \E\norm{\x{i}{t-1}{k} - \x{t-1}}\\
    &\quad + 12\eta^2 \lambda^2 \bar{K}^2 L^2 \sum_{i, K_i \leq \bar{K}} \sum_{k=0}^{K_i-1} \frac{\w_i}{K_i} \E\norm{\x{i}{t-1}{k} - \x{t-1}}\\
    &\quad + 24\eta^2 \lambda^2 \bar{K}^2 L^2 \E\norm{\x{t} - \x{t-1}} + 12\eta^2 \lambda^2 \bar{K}^2 L^2 \E\norm{\x{t} - \x} \\
    &\quad + 3\eta^2\sigma^2\sum_{i=1}^M \w_i^2 K_i + 3\eta^2 \lambda^2 \sigma^2 \sum_{i=1}^M \w_i^2 \bracket{\frac{\bar{K}}{K_i} - 1}^2 K_i\\
    &\quad + 12 \eta^2 \lambda^2 \sigma^2 \sum_{i, K_i > \bar{K}} \frac{\w_i^2 \bar{K}^2}{K_i}\\
    &\quad + \frac{16\eta L^2}{\mu} \sum_{i=1}^M \sum_{k=0}^{K_i-1}\w_i \bracket{\E\norm{\x{i}{t}{k} - \x} + \E\norm{\x{i}{t-1}{k} - \x}}\\
    &\quad + \frac{16\eta \bar{K} L^2}{\mu} \E\norm{\x{t} - \x{t-1}} \\
    &\quad + \frac{16\eta \bar{K} L^2}{\mu} \sum_{i, K_i \leq \bar{K}} \sum_{k=0}^{K_i-1} \frac{\w_i}{K_i} \E\norm{\x{i}{t-1}{k} - \x{t-1}}\\
    &\quad + 2 \eta \bar{K} \bracket{F(\x) - F(\x{t}) + L\E\norm{\x{t} - \x{t-1}}}
\end{align*}
By observation, there are two recursive formulas, i.e., $\E\norm{\x{t} - \x{t-1}}$ and $\sum_{i=1}^M \sum_{k=0}^{K_i-1}\w_i K_i \E\norm{\x{i}{t-1}{k} - \x{t-1}}$, and the coefficients of both of them contain the stepsize $\eta$. To release these formulas, we let the formula on the left-hand side be the formula as follows: 
\begin{align*}
    \mathcal{Y}_{t+1} &= \E\norm{\x{t+1} - \x} + u_1 \E\norm{\x{t+1} - \x{t}}\\
    &\quad + u_2 \sum_{i=1}^M \sum_{k=0}^{K_i-1}\w_i K_i \E\norm{\x{i}{t}{k} - \x{t}}
\end{align*}
where $u_1$ and $u_2$ are the coefficients containing the stepsize of $\eta$ such that the coefficient for $\E\norm{\x{t} - \x{t-1}}$ and $\sum_{i=1}^M \sum_{k=0}^{K_i-1}\w_i K_i \E\norm{\x{i}{t-1}{k} - \x{t-1}}$ on the right hand side become negative and thereby, can be omitted when finding the bound. Also, we notice that the added term includes $\eta^2$ after simplification. Therefore, the formula can be further simplified as: 
\begin{align*}
    \mathcal{Y}_{t+1}&\leq \bracket{1-\frac{\eta\mu\bar{K}}{4}} \mathcal{Y}_t - 2\eta \bar{K} (F(\x{t}) - F(\x) ) \\
    &\quad + 6\eta^2\sigma^2\sum_{i=1}^M \w_i^2 K_i + 6\eta^2 \lambda^2 \sigma^2 \sum_{i=1}^M \w_i^2 \bracket{\frac{\bar{K}}{K_i} - 1}^2 K_i \\
    &\quad + 24 \eta^2 \lambda^2 \sigma^2 \sum_{i, K_i > \bar{K}} \frac{\w_i^2 \bar{K}^2}{K_i}\\
\end{align*}
The rest step is similar to the proof in Theorem \ref{tm:tm1}. Therefore, we can obtain the expected result in Theorem \ref{tm:tm3}. Different from Theorem \ref{tm:tm1}, this theorem release the term of data heterogeneity and therefore, our result can successfully converge to the global optimizer. 

\end{document}